\input harvmac


\def\unlockat{\catcode`\@=11}
\def\lockat{\catcode`\@=12}

\unlockat

\def\newsec#1{\global\advance\secno by1\message{(\the\secno.
 #1)}
\global\subsecno=0\global\subsubsecno=0\eqnres@t\noindent
{\bf\the\secno. #1}
\writetoca{{\secsym} {#1}}\par\nobreak\medskip\nobreak}
\global\newcount\subsecno \global\subsecno=0
\def\subsec#1{\global\advance\subsecno
by1\message{(\secsym\the\subsecno. #1)}
\ifnum\lastpenalty>9000\else\bigbreak\fi\global\subsubsecno=0
\noindent{\it\secsym\the\subsecno. #1}
\writetoca{\string\quad {\secsym\the\subsecno.} {#1}}
\par\nobreak\medskip\nobreak}
\global\newcount\subsubsecno \global\subsubsecno=0
\def\subsubsec#1{\global\advance\subsubsecno by1
\message{(\secsym\the\subsecno.\the\subsubsecno. #1)}
\ifnum\lastpenalty>9000\else\bigbreak\fi
\noindent{\it\secsym\the\subsecno.\the\subsubsecno.} 
{$\underline{\hbox{\it{#1}}}$}
\writetoca
{\string{\secsym\the\subsecno.\the\subsubsecno.} {\it {#1}}}
\par\nobreak\medskip\nobreak}

\def\subsubseclab#1{\DefWarn#1\xdef
#1{\noexpand\hyperref{}{subsubsection}%
{\secsym\the\subsecno.\the\subsubsecno}%
{\secsym\the\subsecno.\the\subsubsecno}}%
\writedef{#1\leftbracket#1}\wrlabeL{#1=#1}}
\lockat

\def\a{\alpha}    \def\b{\beta}       \def\c{\chi}       
\def\d{\delta}
    \def\e{\varepsilon}        

\def\g{\gamma}          \def\k{\kappa}     

\def\L{\Lambda}   \def\m{\mu}                 

\def\vr{\varrho}  \def\o{\omega}           
\def\p{\psi}
      \def\s{\sigma}           

\def\t{\tau}        \def\w{\varphi}    
\def\x{\xi}

\def\CA{{\cal A}}

\def\CG{{\cal G}}
\def\CM{{\cal M}}
\def\CN{{\cal N}}
\def\CE{{\cal E}}

\def\CW{{\cal W}}
 \def\CV{{\cal V}}
%
\font\teneufm=eufm10
\font\seveneufm=eufm7
\font\fiveeufm=eufm5
\newfam\eufmfam
\textfont\eufmfam=\teneufm
\scriptfont\eufmfam=\seveneufm
\scriptscriptfont\eufmfam=\fiveeufm
\def\eufm#1{{\fam\eufmfam\relax#1}}

\font\teneusm=eusm10
\font\seveneusm=eusm7
\font\fiveeusm=eusm5
\newfam\eusmfam
\textfont\eusmfam=\teneusm
\scriptfont\eusmfam=\seveneusm
\scriptscriptfont\eusmfam=\fiveeusm

\font\tenmsx=msam10
\font\sevenmsx=msam7
\font\fivemsx=msam5
\font\tenmsy=msbm10
\font\sevenmsy=msbm7
\font\fivemsy=msbm5
\newfam\msafam
\newfam\msbfam
\textfont\msafam=\tenmsx  \scriptfont\msafam=\sevenmsx
  \scriptscriptfont\msafam=\fivemsx
\textfont\msbfam=\tenmsy  \scriptfont\msbfam=\sevenmsy
  \scriptscriptfont\msbfam=\fivemsy

\def\msbm#1{{\fam\msbfam\relax#1}}
\font\tenbifull=cmmib10 
\font\tenbimed=cmmib10 scaled 800
\font\tenbismall=cmmib10 scaled 666
\textfont9=\tenbifull \scriptfont9=\tenbimed
\scriptscriptfont9=\tenbismall

\def\cmp#1#2#3{Comm.\ Math.\ Phys.\ {{\bf #1}} {(#2)} {#3}}
\def\pl#1#2#3{Phys.\ Lett.\ {{\bf #1}} {(#2)} {#3}}
\def\np#1#2#3{Nucl.\ Phys.\ {{\bf #1}} {(#2)} {#3}}

\def\ijmp#1#2#3{Int.\ J.\ Mod.\ Phys.\ {{\bf #1}} 
{(#2)} {#3}}
\def\mpl#1#2#3{Mod.\ Phys.\ Lett.\ {{\bf #1}} {(#2)} {#3}}
\def\jdg#1#2#3{J.\ Differ.\ Geom.\ {{\bf #1}} {(#2)} {#3}}

\def\top#1#2#3{Topology {{\bf #1}} {(#2)} {#3}}

\def\ap#1#2#3{Ann.\ Phys.\ {{\bf #1}} {(#2)} {#3}}

\def\mm#1#2#3{Manuscripta \ Math.\ {{\bf #1}} {(#2)} {#3}}
\def\ma#1#2#3{Math.\ Ann.\ {{\bf #1}} {(#2)} {#3}}

\def\im#1#2#3{Invent.\ Math.\ {{\bf #1}} {(#2)} {#3}}
\def\plms#1#2#3{Proc.\ London Math.\ Soc.\ {{\bf #1}} 
{(#2)} {#3}}
\def\dmj#1#2#3{Duke Math.\  J.\ {{\bf #1}} {(#2)} {#3}}

\def\jgp#1#2#3{J.\ Geom.\ Phys.\ {{\bf #1}} {(#2)} {#3}}

\def\rd{\partial}

\def\darr#1{\raise1.5ex\hbox{$\leftrightarrow$}
\mkern-16.5mu #1}

\def\Fr#1#2{{#1\over#2}}
\def\tr{\hbox{tr}\,}

\def\roughly#1{\raise.3ex\hbox{$#1$\kern-.75em
\lower1ex\hbox{$\sim$}}}
\def\ato#1{{\buildrel #1\over\longrightarrow}}

\def\pr{\prime}
\def\ppr{{\prime\prime}}
\def\CH{{\cal H}}
\def\bs{{\bf s}}
\def\bbs{{\bf\bar s}}
\def\Dp{\rd_{\!A}}
\def\Dpp{\bar\rd_{\!A}}
\def\ack{\bigbreak\bigskip\bigskip\centerline{
{\bf Acknowledgments}}\nobreak}
\baselineskip=14pt plus 1.2pt minus .6pt
\newskip\normalparskip
\normalparskip = 4pt plus 1.0pt minus .5pt
\parskip = \normalparskip
\parindent= 16pt

\def\lin#1{\noindent {$\underline{\hbox{\it #1}}$} \par}

\def\CN{\eufm{M}}
\def\CV{\eufm{A}}
\font\Titlerm=cmr10 scaled\magstep3
\nopagenumbers
\rightline{ITFA-96-50, hep-th/9612096
}

\vskip .7in

\centerline{\fam0\Titlerm 
Monads and D-instantons
}

\tenpoint\vskip .4in\pageno=0

\centerline{
Jae-Suk Park
\footnote{$^{\dagger}$}{e-mail: park@phys.uva.nl}
}
\vskip .2in
\centerline{\it Instituut voor Theoretische Fysica}
\centerline{\it  Universiteit van Amsterdam}
\centerline{\it Valckenierstraat 65, 1018 XE Amsterdam}

\vskip .4in

\noindent
\abstractfont

Motivated by  twisted ${\cal N}=4$ supersymmetric Yang-Mills
theory in $4$ dimensions,  a natural extension of the 
monad (ADHM) construction relevant to
D-instantons is considered. 
We show  that a family of  Yang-Mills 
instantons can be constructed from  D-instantons. 
We discuss some possible
roles of  reciprocity in D-brane physics. 
We conjecture the existence of universal instantons
together with a generalized Fourier-Nahm transformation
as an unifying framework of D-brane physics.

\tenpoint

\Date{November, 1996}

\lref\DonaldsonA{
S.~K.~Donaldson, 
{\it Instantons and geometrical invariant theory},
\cmp{93}{453}{1984}
}
\lref\DonaldsonB{
S.~K.~Donaldson, 
{\it Nahm's equations and the classification of monopoles},
\cmp{96}{387}{1984}
}
\lref\DonaldsonC{
S.~K.~Donaldson,
{\it Anti-self-dual Yang-Mills connections on complex
algebraic surfaces and stable vector bundles},
\plms{3}{1985}{1}\semi
{\it Infinite determinants, stable bundles and curvature},
\dmj{54}{1987}{231}
}
\lref\DK{
S.~K.~Donaldson and P.B.~Kronheimer,
{\it The geometry of four-manifolds},
Oxford Mathematical Monographs, Clarendon Press, Oxford,
1990
}
\lref\WardA{
R.~S.~Ward,
\pl{61A}{1977}{81}
\semi
M.~F.~Atiyah and R.S.~Ward,
{\it Instantons and algebraic geometry}
\cmp{55}{1977}{117}
}
\lref\HitchinA{
N.~J.~ Hitchin,
{\it Monopoles and geodesics},
\cmp{83}{1982}{579}
}
\lref\HitchinB{
N.~J.~ Hitchin,
{\it Nahm's equations and the classification of
Monopoles},
\cmp{96}{1984}{387}
}
\lref\CGO{
E.~Corrigan and P.~Goddard,
{\it Construction of instanton and monopoles
solutions and reciprocity},
\ap{154}{1984}{253}
}
\lref\NahmA{
W.~Nahm,
{\it Multimonopoles i the ADHM construction,}
in ``Proc.~Symp. on particle physics'',
eds. Z.~Horvath et al., (Vsiegrad, 1981)
\semi
{\it Construction of all self-dual monopoles by the ADHM method},
in ``Monopoles in quantum field theory''.
eds. N.~Craigie et al., (World Scientific, Singapore, 1982)
}
\lref\NahmB{
W.~Nahm,
{\it Self-dual monopoles and calorons,}
in Lecture Notes in Physics, vol.~201,
eds.~G.~Denardo et al., (Springer 1984)
}
\lref\BB{
P.~J.~Braam and P.~ van Baal,
{\it Nahm's transformation for instantons},
\cmp{122}{1989}{267}
}
\lref\ADHM{
M.~Atiyah, V.~Drinfeld, N.~Hitchin and Yu.~Manin,
{\it Construction of instantons},
\pl{65A}{1978}{185}
}
\lref\VW{
C.~Vafa and E.~Witten,
{\it A strong coupling test of $S$-duality},
\np{B 431}{1994}{3}.
}
\lref\SWa{
N.~Seiberg and E.~Witten,
{\it Electric-magnetic duality, monopole condensation, 
and confinement
in $N=2$ supersymmetric Yang-Mills theory},
\np{B 426}{1994}{19}; 
{\it Monopoles, duality, and chiral symmetry breaking in 
$N=2$ supersymmetric QCD},
\np{B 431}{1994}{484}.
}
\lref\Barth{
W.~Barth,
{\it Some properties of stable rank-2 vector bundles on 
$P_n$},
\im{42}{1977}{63}\semi
W.~Barth and K.~Hulek,
{\it Monads and moduli of vector bundles},
\mm{25}{1978}{125}.
}
\lref\Horrocks{
G.~Horrocks,
{\it Vector bundles on the punctured spectrum of 
a local ring},
\plms{14}{1964}{684}
}

\lref\Nahm{
W.~Nahm,
{\it Multimonopoles i the ADHM construction,}
in ``Proc.~Symp. on particle physics'',
eds. Z.~Horvath et al., (Vsiegrad, 1981);
{\it Construction of all self-dual monopoles by 
the ADHM method},
in ``Monopoles in quantum field theory''.
eds. N.~Craigie et al., 
(World Scientific, Singapore, 1982);
{\it Self-dual monopoles and calorons,}
in Lecture Notes in Physics, vol.~201,
eds.~G.~Denardo et al., (Springer 1984)
}
\lref\DOM{
M.~R.~ Douglas, G.~Moore
{\it D-branes, Quivers, and ALE Instantons},
hep-th/9603167 
}
\lref\Dg{
M.~R. Douglas
{\it Branes within Branes},
hep-th/9512077.
\semi
{\it Gauge Fields and D-branes},
hep-th/9604198.
}
\lref\Pol{
J.~Polchinski,
{\it Dirichlet-Branes and Ramond-Ramond Charges},
hep-th/9510017 
\semi
J.~Polchinski, S.~Chaudhuri, C.V.~Johnson,
{\it Notes on D-Branes}
hep-th/9602052 
}
\lref\WittenA{
E.~Witten,
{\it Twistor-like transform in ten dimensions},
\npb{266}{1986}{245}
}
\lref\WittenB{
E.~Witten,
{\it Monopoles and four-manifolds},
Math.~Research Lett.~{\bf 1} (1994) 769.
}
\lref\WittenC{
E.~Witten,
{\it Sigma models and the ADHM construction of 
instantons},
hep-th/9410052 
}
\lref\WittenD{
E.~Witten
{\it Small instantons in string theory},
hep-th/9511030
}
\lref\WittenE{
E.~Witten,
{\it Bound states of strings and $p$-branes},
hep-th/9510135
}
\lref\WittenF{
E.~Witten,
{\it Topological quantum field theory}
\cmp{117}{353}{386}
}
\lref\WittenG{
E.~Witten,
{\it Two dimensional gauge theories revisited},
\jgp{9}{1992}{303}
}
\lref\DH{
J.~J.~Duistermaat and G.~J.~Heckman,
{\it On the variation in the cohomology of the 
symplectic form of the reduced phase space},
\im{69}{1982}{259}; Addendum, \im{72}{1983}{153}.
}
\lref\Wu{
S.~Wu,
{\it An integration formula for the square of momentum
maps of circle actions},
\lmp{29}{1993}{311}
}
\lref\GS{
V.~Guillemin and S.~Sternberg,
{\it Birational equivalence in symplectic category},
\im{97}{1989}{485}.
}
\lref\Kirwan{F.~Kirwan, Cohomology of quotients in symplectic
and algebraic geometry (Princeton Univ.~Press, 1987).
}
\lref\DLP{
J.~Dai, R.~G.~Leigh, J.~Polchinski,
{\it 
New connections between string theories},
\mpl{A4}{1989}{2073}
}
\lref\BPST{
A.~Belavin, A.~Polyakov, S.~Schwartz and Y.~Tyupkin,
{\it Psseudoparticle solutions of the Yang-Mills
equations
}
\pl{B59}{1975}{85}
}
\lref\Atiyah{
M.~F.~Atiyah,
{\it Geometry of Yang-Mills fields,}
Scuola Normale Superiore,
Pisa, 1979.
}
\lref\Shenk{
H.~Schenk,
{\it On a generalized Fourier transformation of instantons
over flat tori,}
\cmp{116}{1988}{183}
}
\lref\Macio{
A.~Maciocia,
{\it Metrics on the moduli spaces of instantons
over Euclidean $4$-space},
\cmp{135}{1991}{467}
}
\lref\MaciaB{
A.~Maciocia,
{\it The determinant line bundle over moduli spaces
of instantons on Abelian surface},
\mz{217}{1994}{317}
}
\lref\BW{
E.~Br\'{e}zin and S.~R.~Wadia, eds.,
{\it
The large $N$ expansion in quantum field theory
and statistical physics; from spin systems to
$2$-dimensional gravity},
World Scientific, (Singapore, 1993).
}
\lref\KN{
P.~B.~Kronheimer and H.~Nakajima,
{\it Yang-Mills instantons on ALE gravitational instantons},
\ma{288}{1990}{263}
}
\lref\Mukai{
S.~Mukai, {\it Dulaity between $D(X)$ and $D(\hat X)$,
with application to Picard sheaves},
Nagoya Math.~J.~{\bf 81} (1981) 153.
}
\lref\DED{
D.-E.~Diaconescu, 
{\it  D-branes, Monopoles and Nahm Equations},
hep-th/9608163
}
\lref\DM{
R. Dijkgraaf, G. Moore,
{\it Balanced Topological Field Theories},
hep-th/9608169 
}

\lref\WittenZ{
E.~Witten,
{\it Supersymmetric Yang-Mills theory on a four-manifold},
J.~Math.~Phys.~{\bf 35} (1994) 5101,
hep-th/9403195 
{\it 
Monopoles and four-manifolds},
hep-th/941110
}
\lref\WittenT{
E.~Witten
{\it Two dimensional gauge theories revisited}, 
J.~Geom.~Phys.~ {\bf 9} (1992) 303
hep-th/9204083 
}

\lref\Witten{ 
E.~Witten,
{\it Topological quantum field theory},
\cmp{117}{1988}{353};
{\it Introduction to cohomological field theories},
\ijmp{A6}{1991}{2273}
}
\lref\DPS{
R.~Dijkgraaf, J.-S.~Park and B.~J.~Schroers,
{\it to appear}.
}
\lref\HPB{
S.~Hyun and J.-S.~Park,
{\it Holomorphic Yang-Mills theory and the
variation of the Donaldson polynomials},
hep-th/9503092.
}
\lref\HYM{
J.-S.~Park,
{\it N=2 topological Yang-Mills theory
on compact K\"{a}hler surfaces},
\cmp{163}{1994}{113};
{\it Holomorphic Yang-Mills theory on compact
K\"{a}hler manifolds};
\np{B423}{1994}{559}
}
\lref\HPA{
S.~Hyun and J.-S.~Park,
{\it N=2 topological Yang-Mills theories
and Donaldson's polynomials},
J.~Geom.~Phys.~{\bf 20} (1996) 31, 
hep-th/9404009
}
\lref\Donaldson{
S.~K.~Donaldson,
{\it Polynomial invariants for smooth $4$-manifolds},
\top{29}{1990}{257}
}
\lref\tHooft{
G.~'t Hooft,
{\it A planar diagram theory for strong interactions},
\np{B72}{1974}{461}
}
\lref\IZ{
C.~Itzykson and J.~B.~Zuber,
{\it The planar approximation II},
J.~Math.~Phys.~{\bf 21} (1980) 411
}
\lref\Doug{
M.~R.~Douglas,
{\it String in less than one dimension and the generalized
KdV hierarchies},
\pl{B238}{1990}{176}
}
\lref\Nakajima{
H.~Nakajima,
{\it Instantons on ALE spaces, quiver varieties,
and Kac-Moody algebra},
\dmj{76}{1994}{365};
{\it Homology of moduli spaces of instantons on
ALE spaces. I},
\jdg{40}{1994}{105};
{\it Instantons and affine Lie algebras},
alg-geom/9510003
}

\lref\NakajimaB{
H.~Nakajima,
{\it Heisenberg algebra and Hilbert schemes of
points on projective surfaces},
alg-geom/9507012
}
\lref\NakajimaC{
H.~Nakajima,
{\it Gauge theory on resolutions of simple singularities and 
simple Lie algebras},
Int.~Math.~Res.~Not.~{\bf 2} (1994) 61;
{\it  Gauge theory on resolutions of simple singularities and 
affine Lie algebras};
{\it Resolutions of moduli spaces of ideal instantons
on $\msbm{R}^4$},
in ``Topology, geometry and field theory'',
World Scientific, (Singapore, 1994).
}
\lref\NakajimaD{
H.~Nakajima,
{\it Monopoles and Nahm's equations},
in Einstein  metric and Yang-Mills connections,
eds.~T.~Mabuchi and S.~Mukai,
(Marcel Dekker, Inc.~1993)
}
\lref\NakajimaP{
H.~Nakajima, {\it private communication}.
}
\lref\Gott{
L.~G\"{o}ttsche,
{\it The Betti numbers of the Hilbert scheme of points
on a smooth projective surface},
\ma{286}{1990}{193}
}
\lref\Park{
J.-S.~Park, {\it Monadic Yang-Mills theory},
in preparation.
}
\lref\Vafa{
C.~Vafa,
{\it Instantons on D-branes},
hep-th/9512078
}
\lref\DVV{
R.~Dijkgraaf, E.~Verlinde and H.~Verlinde,
{\it Counting Dyons in {\cal N}=4 String Theory},
hep-th/9607026 
\semi
R.~Dijkgraaf, G.~Moore, E.~Verlinde and H.~Verlinde,
{\it Elliptic Genera of Symmetric Products and 
Second Quantized Strings},
hep-th/9608096 
}
\lref\Picken{
R.F.~Picken,
{\it The Duistermaat-Heckman integration
formula on flag manifolds},
J. Math. Phys. {\bf 31} (1990) 616.
}
\lref\JK{
L.C.~Jeffrey and F.C.~Kirwan,
{\it Localization for non-abelian group actions},
preprint alg-geom/9307005.
}
\lref\HL{
Y.~Hu and W.-P.~Li,
Variation of the Gieseker and Uhlenbeck compactifications,
preprint alg-geom/9409003.
}
\lref\DoH{
I.~V.~Dolgachev and Y.~Hu,
Variation of geometrical invariant theory Quotients,
preprint.
}
\lref\Donald{
S.~K.\ Donaldson, 
{\it Polynomial invariants for smooth $4$-manifolds},
\top{29}{1990}{257}.
}
\lref\VafaF{
C.~Vafa,
{\it Evidence for F-theory},
hep-th/9602022.
}
\lref\BSV{
M.~Bershadsky, V.~Sadov and C.~Vafa,
{\it
D-branes and topological field theories},
hep-th/9511222
}
\lref\ST{A.~Strominger, 
{\it Heterotic solitons},
\np{B343}{1990}{167}\semi
C.G.~Callan, Jr., J.A.~Harvey and A.~Strominger,
{\it Worldbrane actions for string solitons},
\np{B367}{1991}{60}
}
\lref\MAT{
P. Townsend, 
{\it D-branes from M-branes}, 
\pl{B373}{1996}{68}\semi
T.~Banks, W.~Fischler, S.~H.~Shenker and  L.~Susskind,
{\it  M theory as a matrix model: a conjecture},
hep-th/9610043 
\semi
M.~Berkooz and M.~Douglas,
hep-th/9610236
\semi
V.~Periwal,
{\it Matrics on a point as the theory of everything},
hep-th/9611103
}
\lref\MT{
M.~J.~Duff, P.~Howe, T. Inami, K.S. Stelle, 
{\it Superstrings in D=10 
{}from supermembranes in D=11}, 
\pl{B191}{1987}{70}\semi
M.~J.~ Duff, J. X. Lu, 
{\it Duality rotations in membrane theory}, 
\np{B347}{1990}{394} \semi 
M.~J.~ Duff, R. Minasian, James T. Liu, 
{\it leven-dimens ional origin of 
string/string duality: a one-loop test}, 
\np{B452}{1995}{261} \semi
C.~ Hull and P.~ K.~ Townsend, 
{\it Unity of superstring dualities},
\np{ B438}{1995}{109},
\semi
P.~ K.~ Townsend, 
{\it The Eleven-Dimensional Supermembrane Revisited},
\pl{B350}{1995}{184};
{\it String-membrane duality in seven dimensions}, 
\pl{ 354B} {1995}{ 247}, \semi
C.~ Hull and P.~ K.~ Townsend,
{\it Enhanced gauge symmetries in superstring theories}, 
\np{B451}{1995}{525},  \semi
E.~ Witten,
{\it String Theory Dynamics in Various Dimensions},
\np{B443}{1995}{85} 
}
\lref\GR{
I.~Grojnowski, 
{\it
Instantons and affine algebras I: the Hilbert scheme
and vertex operators},
alg-geom/9506020.
}
\lref\ZB{
U.~H.~ Danielsson, G.~ Ferretti and B.~ Sundborg,
{\it D-particle Dynamics and Bound States}, 
hep-th/9603081\semi 
D.~ Kabat and P.~ Pouliot,
{\it A
Comment on Zero-Brane Quantum Mechanics}, 
hep-th/9603127
\semi M.~R.~ Douglas, D.~ Kabat, P.~ Pouliot and S.~ Shenker, 
{\it D-Branes and Short Distance 
in String Theory}, hep-th/9608024 .
} 
\lref\LNN{
H.~Nakajima,
{\it Lectures on Hilbert schemes of points on
surfaces},
Lecture Notes given at Dep. of Math. Sci., Univ. of Tokyo,
to appear.
}
\lref\WHN{
B.~de Wit, J.~Hoppe and H.~Nicolai,
\np{B305}{1988}{545}; 
\np{B320}{1989}{135}.
}
\lref\FU{
C.~Johnson and R.~Myers,
{\it Aspects of type IIB theory on ALE spaces},
hep-th/9610140.
}
\lref\Taylor{
W.~Taylor IV,
{\it D-brane field theory on compact spaces},
hep-th/9611042
}
\lref\KHT{
K.~Furuuchi, H.~Kunitomo and T.~Nakatsu,
{\it Topological field theory and second-quantized
five-branes}, hep-th/9610016.
}
\lref\DT{
S.~Donaldson, {\it Lectures given at I.~Newton
Inst.}, Nov.~1996.
}
\lref\REV{
M.R.~Douglas,
{\it Superstring dualities, Dirichlet branes
and the samll scale structure of space},
hep-th/9610041.
}

\newsec{Introduction}

Shortly after the discovery of the Yang-Mills 
instantons \BPST,
Atiyah, Dringfeld, Hitchin and Manin (ADHM) 
found of a way of constructing, in principle, 
all instantons for any classical group \ADHM.
The ADHM construction relates the moduli space of  
anti-self-dual  connections on  $\msbm{R}^4$ to the 
space of monads via the Ward correspondence \WardA.
Subsequently, Donaldson reformulated the ADHM construction 
in the language holomorphic vector bundle on 
$\msbm{CP}^2$ \DonaldsonA. 
The ADHM construction can be generalized
to the Fourier-Nahm-Mukai transformation \Nahm\Mukai.
For any sub group $H$ of the group of isometries of $\msbm{R}^4$,
this transformation gives an isometry  
between the moduli space  
of $H$-invariant instantons on $\msbm{R}^4$ 
and the moduli space of  $H^*$-invariant instantons
on $\msbm{R}^{*4}$ \Nahm\BB\Macio\NakajimaD. 
This isometry and its miraculous properties, 
originally pointed out by Corrigan and Goddard, is called
the reciprocity \CGO.

In the recent developments of string theory
D-branes play crucial roles \Pol\REV.
The low energy theory on the D $p$-branes
world volume is the dimensional reduction of
$d=10$ ${\cal N} =1$ super-Yang-Mills theory down to
$p+1$ dimensions \DLP\WittenE.
The type IIB string has
$p=-1$ D-branes or D-instantons whose
low energy  description has an intriguing
similarity with the ADHM construction 
of Yang-Mills instanton \WittenE.

Motivated by some properties of a
${\cal N}=4$ supersymmetric Yang-Mills
theory in $4$-dimensions \VW\DPS, we will consider
a simple extension of the monad (ADHM) construction.
We will show that such an extension is 
closely related to D-instantons, 
which are, in turn, capable of being frozen
on parallel $4$-dimensional Euclidian spaces 
to become a family of genuine Yang-Mills instantons
of different gauge groups and  instanton numbers.
Equivalently, we get a family of the ADHM constructions.
This identification suggest that reciprocity (maybe in
a suitably generalized form) may have a significant
role in the D-brane physics.
Indeed, Witten's description of Dirichlet 
p-branes  is very suggestive of such a role \WittenE. 
The various D-branes
may be related to certain universal instantons
invariant under suitable isometry $H$. 
The usual description of
D-branes may be obtained by  doing a sort of 
Fourier-Nahm-Mukai transformation. 

We should mention several earlier papers relating
D-branes and Yang-Mills instantons.
Witten showed that
the type I D $5$-brane is a point-like 
instanton \WittenD\ using his ADHM linear 
sigma model \WittenC. 
Douglas extend the picture and produced the  instantons
using certain D-branes as the probes \Dg.
Douglas and Moore reproduced
results of Kronheimer and Nakajima on Yang-Mills 
instantons in ALE space \KN. They also suggested a
relation between $T$-duality and reciprocity \DOM. 
Related issues have also studied in \FU\KHT.
Diaconescu's paper shows some interesting
relation between the ADHMN construction of monopole
D-branes  \DED.

In the above mentioned cases
the Yang-Mills instantons are constructed,
for example, via the world-sheet theory of a D 1-brane
moving in the background of a system of D 5-branes
and 9-branes, depending on the rank of the gauge group 
and on the instanton numbers. 
Our construction will be quite different from the above
approach. We will show a direct relation between
D-instantons and Yang-Mills instantons.
It will lead us to suggest
the possibility of reversing the usual program: 
D-brane physics may be described by certain universal 
monads 
in unifying way.

This paper is organized as follows,
In Sect.~2, we review the ADHM construction.
We first recall the result of the construction
and review the reformulation of Donaldson
who used monads on $\msbm{CP}^2$ and
geometric invariant theory.
In Sect.~3, we present a simple extension
of Donaldson's construction. In this extended
form there is no distinction between the
rank of the gauge group and the instanton number.
We use the natural $S^1$ action on the space of
extended monads and study its fixed points.
The usual ADHM construction can be recovered
as a special limit. 
In Sect.~4,
we  show that the extended monads
construction can be obtained by dimensional
reduction of a ${\cal N}=2$ preserving perturbation
of a ${\cal N}=4$ super-Yang-Mills theory in $4$-dimensions.
We show that a further perturbation
to a ${\cal N} =1$ theory leads to a resolution
of singularities. 
We also formulate natural matrix models related to
monads.
In Sect.~5, 
we  study the relation between the extended
monads, D-instantons and Yang-Mills instantons. 
Finally, some implications for D-brane physics
are discussed.

\newsec{Preliminary}

This section is devoted to a brief review of the
ADHM construction and related topics.
The contents of this section are
classical and several good reviews exist \Atiyah\CGO\DK.  
The main purpose is to establish our notations
and provided some background for the discussions
in the later sections.  Most of the material here  
is based on \DK\ and \DonaldsonA, to which we refer
for details. See also \LNN\ for a self-contained
survey and many other issues relevant to this paper.

The ADHM construction is based on the twistor method
via the Ward correspondence \WardA\ and the monads of 
Horrocks \Horrocks\ and Barth \Barth\ on algebraic 
bundles over $\msbm{CP}^3$. 
Donaldson reformulated the ADHM construction
by identifying instantons on $\msbm{R}^4$ with 
stable holomorphic
bundles over $\msbm{CP}^2$ trivial along the line
at infinity \DonaldsonA\DonaldsonC.  
In both constructions, the notion of monads plays the
central role. Most of our paper will be
algebraic, so we will recall Nahm's approach
only when it is necessary.

\subsec{The ADHM constructions}

For our purpose it is sufficient to
recall the end results (Ch.~(3.3) in \DK).
Consider $SU(n)$ instantons with instanton number
$k$. The ADHM data is given by

(i) A $k$-dimensional complex vector space $\CH$ and
a Hermitian metric and self-adjoint linear
maps $T_i:\CH \rightarrow \CH$ with $i=1,2,3,4$.

(ii) An $n$-dimensional complex Hermitian 
vector space $E_\infty$ with determinant form,
and a linear map $P: E_\infty\rightarrow \CH\otimes S^+$, 
where $S^+$ is the positive two-dimensional
spin space of $R^4$.

(iii) There is surjective linear map $R_x$ for each 
$x\in R^4$
$R_x : \CH\otimes S^-\oplus E_\infty \rightarrow\CH\otimes
S^+$ by
\eqn\aba{
R_x =\left(\sum_{i=0}^3(T_i - x_i I)\otimes \gamma(e_i)^*
\right)\oplus P,
}
where $\g(e_i)^*:S^{-}\rightarrow S^+$
are the adjoints of the maps defining the spin structure.
This is the non-degeneracy condition.

Now the ADHM data describe the moduli space of instantons
by the solutions, up to gauge degree of freedom,
of the ADHM equations;
\eqn\abb{\eqalign{
[T_1,T_2] + [T_3,T_4]= PP^*_I,
\cr
[T_1,T_3] + [T_4,T_2]= PP^*_J,
\cr
[T_1,T_4] + [T_2,T_3]= PP^*_K,
}
}
The ADHM data $(T_i,P)$
is equivalent to the data $(T^\pr_i,P^\pr)$
if
\eqn\symm{
T^\pr_i = v T_i v^{-1},\qquad  P^\pr= v P u^{-1},
}
for $v$ in $U(k)$ and $u$ in $SU(n)$

{}From the ADHM data $(T,P)$, one can reconstruct
the explicit connection $A(T,P)$ by the standard
way of introducing a connection via a projection
onto a sub-bundle embedded in a trivial bundle
with fibre $\CH\oplus E_{\infty}$
(for reviews see \Atiyah\DK).
Here the sub-bundle  is defined by
the kernel $E_x$ of the linear map $R_x$ \aba.
The ADHM equation \abb\  ensures
that the connection obtained in this way
is anti-self-dual (ASD).

The non-degeneracy condition means the kernel
is precisely $n$-dimensional so that we have
a rank $n$ non-trivial bundle $E$ (with fibre
$E_x$) over $\msbm{R}^4$.\foot{The $n$ dimensional
space $E_\infty$ becomes the fiber of $E$ at
infinity. In the approach of Nahm $E_\infty$
and the associated map $P$ originated from
the boundary condition at 
infinity of the adjoint Dirac 
equation in the instanton background.
Sometimes the term $P$ is referred to as the
source term which spoils the anti-self-duality
of the $T_i$ matrices \CG\Nahm. 
The source term arises when we are dealing
with non-compact manifold like $\msbm{R}^4$.
}

It is convenient
to introduce the complex variables;
\eqn\aga{
\t_1 = T_1 + i T_2,\qquad \t_2=T_3 + i T_4.
}
We also decompose $S^+$ in two pieces such that
\eqn\agb{
P=a^* \otimes <1> + b\otimes <\theta>,
}
where $a:\CH\rightarrow E_\infty$, 
$b:E_\infty\rightarrow \CH$.
Now the ADHM equation becomes
\eqn\abg{\eqalign{
[\t_1,\t_2] +b a=0,\cr
[\t_1,\t^*_1] + [\t_2,\t_2^*] +b b^* - a^*\!a=0,
}
}
where $\t_1,\t_2$ are $k\times k$ matrices,
$b$ and $a$ are $k\times n$ and $n\times k$
matrices.
The ADHM data can be expressed by the following
figure \LNN;

\input epsf 

\bigskip
     \epsfxsize=4truecm
\centerline{
\epsfbox{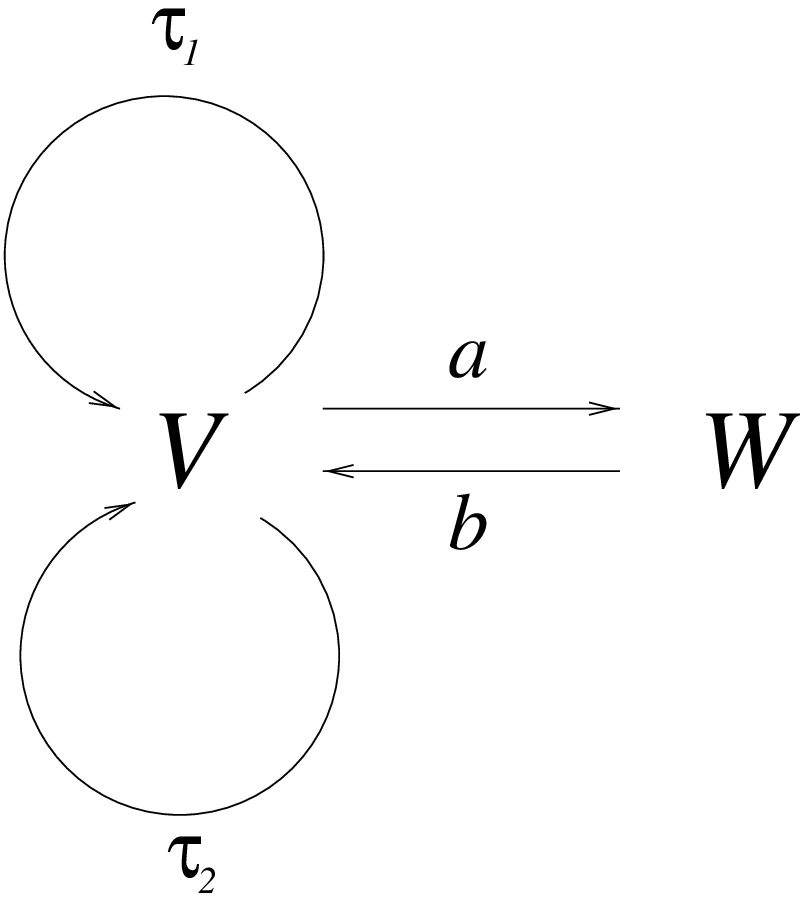}
}
\medskip
{\footskip14pt plus 1pt minus 1pt \footnotefont
{\bf Fig.~1.} 
The ADHM data for $k$ $SU(n)$ instantons: 
$V=\msbm{C}^k$, $W=\msbm{C}^n$, $\t_1,\t_2 \in End(V)$,
$a\in Hom(V,W)$ and $b\in Hom(W,V)$.
}
 \bigskip

\subsec{The Approach of Donaldson}

Donaldson reformulated the ADHM construction 
in terms of geometric invariant theory \DonaldsonA.
Let $E$ be an $SU(n)$ bundle over $\msbm{CP}^2$
and consider the moduli space of ASD connections
on it.  Donaldson identified the instanton moduli space
with the moduli space of stable holomorphic
$SL(n,\msbm{C})$ bundles $\CE$ which are topologically 
equivalent to $E$ \DonaldsonC.   
Now $\msbm{R}^4$ can be naturally regarded as 
$\msbm{C}^2$ by picking a complex structure $I$
with $I^2=-1$.
Donaldson  regarded the flat metric
on $\msbm{C}^2$ as a metric on $\msbm{CP}^2$
singular along a (complex) line $\ell_\infty$
at infinity. Then an instanton on $\msbm{R}^4$
can be identified with an equivalence class of 
stable holomorphic
bundles on $\msbm{CP}^2$ which is trivial
precisely along the line $\ell_\infty$.
So the remaining task is to find such a stable bundle.

This can be accomplished by
introducing a {\it monad}.
Consider homogeneous coordinates
$[X]= (x,y,z)$ for $\msbm{CP}^2$ with $z=0$
defining the line $\ell_\infty$ at infinite.
Let $H,K,L$ be complex vector spaces of dimensions
$k,2k+n,k$, respectively. A monad is linear
maps for each $X\in \msbm{C}^3$;
\eqn\seqe{
H \ato{A_X} K \ato{B_X} L,
}
depending linearly on $X$:
\eqn\caa{
A_X = A_x x + A_y y + A_z z,
\qquad
B_X = B_x x + B_y y + B_z z,
}
such that $B_X A_X: H \rightarrow L$ is zero.
This gives a set of six equations.
A monad is non-degenerated if for each
non-zero $X$, $A_X$ is injective and $B_X$
is surjective. This corresponds to the 
non-degeneracy condition (Sect.~$2.1.$ (iii))
Then a monad defines
a complex and a holomorphic vector bundle
over $\msbm{CP}^2$ whose fiber at the point $X$
is the vector space $Ker B_X/Im A_X$.
The bundles trivial on a fixed line, say $z=0$,
can be constructed by purely algebraic
considerations.
The five conditions
\eqn\cab{
B_x A_x = B_y A_y =0,\qquad
B_xA_y+B_yA_x = B_zA_y+B_yA_z=B_zA_x + B_xA_z=0,
}
can be solved as follows:
\eqn\cac{
B_z=(\matrix{-\t_2 &\t_1 &b}),\qquad 
A_z=\left(\matrix{\t_1\cr \t_2\cr a}\right),
}
where $\t_1,\t_2$ are $k\times k$ matrices
and $b$ and $a$ are $k\times n$ and
$n\times k$ matrices.
We have a remaining condition:
\eqn\cad{
B_x A_x =0 \Longrightarrow [\t_1,\t_2] + b a=0.
}
This is the first equation of ADHM in \abg.

Let $\CW_{k,n}$ be the space of quadruples 
$(\t_1,\t_2,a,b)$. The space $\overline\CW^{1,1}_{k,n}$ of
monads is defined by \cad. Let
$\CW^{1,1}_{k,n}$ be the space of non-degenerated
monads.
The space $\CW^{1,1}_{k,n}$ can be identified with
the space $\CA^{1,1}_{n,k}$ of framed holomorphic
connections of a $SU(n)$ bundle with instanton number 
$k$.\foot{A connection $A$ is called framed if it
tends to a pure gauge at infinity together with
a trivialization at infinity. Throughout
this paper, we will always consider the framed
connections.}
The non-degeneracy condition is identical
to the stability of holomorphic vector bundles.
On the space of quadruples $(\t_1,\t_2,a,b)$,
we have the natural $GL(k,\msbm{C})$ action;
\eqn\nglk{
(\t_1,\t_2,a,b)\rightarrow 
(p\t_1 p^{-1},p\t_2 p^{-1},a p^{-1},pb),
\qquad p\in GL(k,\msbm{C}).
} 
A fundamental theorem of Donaldson says that
there is one-to-one correspondence
between the quotient space 
$\hat{\CM}_{k,n}=\CW^{1,1}_{k,n}/GL(k,\msbm{C})$ 
and the quotient space$(\CA^{1,1})^s_{n,k}/\CG^{\msbm{C}}$
where $\CG^{\msbm{C}}$ is the complexification
of the infinite-dimensional group $\CG$ of
the gauge transformation. 
The latter space can be identified with the moduli space
$\CM_{n,k}$ of framed ASD connections of 
gauge group $SU(n)$ with the instanton number $k$. 
To get the moduli space of unframed
instantons, we need to consider $SL(n,\msbm{C})$
action together with \nglk:
\eqn\rrr{
(\t_1,\t_2,a,b)\rightarrow 
(p\t_1 p^{-1},p\t_2 p^{-1},q a p^{-1},p b q^{-1}),
\qquad p\in GL(k,\msbm{C}),\quad q \in SL(n,\msbm{C}).
}

The remaining step is to identify 
the equivalence class of stable orbits
of the natural $GL(k,\msbm{C})$ action (the GIT quotient).
This can be done by using the relation 
between GIT quotient and symplectic quotients.
We have a natural Hamiltonian action $U(k)$ on
the quadruples $(\t_1,\t_2,a,b)$ satisfying
\cad\ with momentum map (Hamiltonian) $\m$ 
given by
\eqn\cae{
\m_1(\t_1,\t_2,a,b) 
  = [\t_1,\t_1^*] +[\t_2,\t_2^*] +b b^* - a^*a.
}
This is the second equation of ADHM in \abg.
Now we can form the symplectic
quotient $(\m^{-1}(0)\cap \CW^{1,1}_{k,n})/U(k))$,
which is identical to the GIT quotient $\hat{\CM}_{k,n}$.
The ADHM theorem says it is diffeomorphic to
the  moduli space 
$\CM_{n,k}$ of framed instantons.
There is a stronger theorem which says that the above
two spaces are isometric as hyperK\"{a}hler
manifolds \Macio. One may also view the
two equations \cad\ and \cae\ as the
Hyperk\"{a}hler momentum maps associated with
the hyperk\"{a}hler structure $I,J,K$ of $\msbm{R}^4$;
\eqn\hyperk{
[\t_1,\t_2] + b a = \m_{2} + i\m_3 \equiv \m_{\msbm{C}},
\qquad
[\t_1,\t_1^*] +[\t_2,\t_2^*] +b b^* - a^*a =
\m_{1} \equiv \m_{\msbm{R}}.
}

The  non-degeneracy condition corresponds to
the stability.
If we remove the non-degeneracy condition
the  quotient space will contain semi-stable
orbits. 
This gives a natural completion $\overline{\hat
\CM}_{k,r}$ of the moduli space $\hat \CM_{k,r}$.
This property will be
important in this paper.
If one of the elements of the quadruples is not
stable, $\t_1$ and $\t_2$ can be block
diagonalized
\eqn\jui{
\eqalign{
\t_1 =\left(
\matrix{\t_1^\pr& 0\cr
0& diag(\x^1_{1},\ldots,\x^\ell_{1})}\right),\cr
\t_2 =\left(
\matrix{\t_2^\pr &0\cr
0&  diag(\x^1_{2},\ldots,\x^\ell_{2})}\right),\cr
b = (\matrix{b^\pr&0}),\qquad 
a =\left(\matrix{a^\pr\cr 0}\right)
}
}
where $\t_1^\pr, \t_2^\pr$ are $(k-\ell)\times (k-\ell)$
matrices, $b^\pr$ and $a^\pr$ are $(k-\ell)\times n$
and $n\times (k-\ell)$ matrices.
Then the equation describes $SU(n)$ instantons
of the instanton number $(k-\ell)$ together
with $\ell$  point-like instantons.
One interprets $(\x_1^i, \x_2^i)$ as the complex
coordinates of the position
of the $i$-th point-like instanton.
So one gets 
\eqn\juj{
\overline{\hat\CM}_{k,n}
=\hat\CM_{k,n}\cup \hat\CM_{k-1,n}\times \msbm{C}^2
\cup \ldots 
\cup \hat\CM_{k-\ell,n}\times S^{k-\ell}(\msbm{C}^2)
\ldots\cup S^{k}(\msbm{C}^2),
}
where $S^\ell(X)$ denotes the $\ell$-th symmetric
product of $X$ - the configuration space
of $\ell$  non-interacting bosonic
particles on $X$.
Maciocia showed that this is isometric
to the Donaldson-Uhlenbeck  
compactification
\eqn\ducom{
\overline\CM_{n,k} =\bigcup_{\ell=0}^{k}
\CM_{n,k-\ell}\times S^\ell(\msbm{C}^2),
}
of $\CM_{n,k}$ \Macio.
Clearly the space $\overline{\hat\CM}_{k,n}$ contains
singularities because of the $S^\ell(\msbm{C}^2)$.
 
Through out this paper, we will not impose the
non-degeneracy condition unless specified.
For example,
\eqn\notaa{
\overline{\hat\CM}_{k,n} = 
\left(\m^{-1}_{\msbm{C}}(0)\cap 
\m^{-1}_{\msbm{R}}(0)\right)/U(k).
}
Because of the isometry between the
two moduli spaces $\overline{\hat\CM}_{k,n}$
and $\overline\CM_{n,k}$, 
we will often not
distinguish them.

\newsec{Extension}

\subsec{An Extension of Monads}

Now we consider a simple modification
of the monad construction 
in the previous subsection. 
To begin with we consider monads for
$SU(2N)$ instantons with instanton number $N$.
We have three complex vector spaces $H,K,L$ of dimensions
$N,4N,N$, respectively.
The five conditions \cab\
can be  solved via
\eqn\cac{
\eqalign{
B_x=(\matrix{0&1&0&0}),\cr
A_x=\left(\matrix{1\cr 0\cr 0\cr 0}\right),
}\qquad
\eqalign{
B_y=(\matrix{-1&0&0&0}),\cr
A_y=\left(\matrix{0\cr 1\cr 0\cr 0}\right),
}\qquad
\eqalign{
B_z=(\matrix{-t_2&t_1&-t_4&t_3}),\cr
A_z=\left(\matrix{t_1\cr t_2\cr t_3\cr t_4}\right),
}
}
where $t_1,t_2,t_3,t_4$ are $N\times N$ matrices.
Note that $B = A^t (J\oplus J)$ 
where $J=\left(\matrix{0 & 1 \cr -1&0}\right)$,
in the notation of block matrix.  
The remaining condition $B_x A_x=0$ 
is identical to 
\eqn\cad{
B_x A_x =0 \Longrightarrow [t_1,t_2] + [t_3,t_4]=0.
}
We may put $\left(\matrix{-t_4 & t_3}\right)$
into a $N\times 2N$ matrix, which we call $b$. 
In the standard interpretation,
we are considering a special subset of monads 
for $SU(2N)$ and $N$ instantons.
In \cac,  we fixed a preferred basis for
the  $N\times 2N$ matrix $b$ and $2N\times N$ matrix $a$;
\eqn\exg{
b = a^t J \Longleftrightarrow 
(\matrix{-t_4&t_3})
= (\matrix{t_3 &t_4})\left(\matrix{0 & 1 \cr -1&0}\right).
}
Our preferred choice of the basis is invariant under
the $GL(N,\msbm{C})$ action (for example,
the left action on $b$). 
In contrast to the usual $SU(2N)$ case, our basis
is not invariant under the $SL(2N,\msbm{C})$ action (
for example, the right
action on $b$) but invariant under
$GL(N,\msbm{C})\oplus GL(N,\msbm{C})$ action.
Now we formulate the following $GL(N,\msbm{C})$
action on the  quadruples  $(t_1,t_2,t_3,t_4)$
\eqn\idt{
(t_1,t_2,t_3,t_4) \rightarrow 
(pt_1p^{-1},pt_2p^{-1},pt_3p^{-1},pt_4p^{-1}),
\qquad p\in GL(N,\msbm{C}).
}
This should be compared with \nglk\ and \rrr.
Here we put $(t_1,t_2)$ and $(t_3,t_4)$ on 
equal footing.

Let $\CV_N$ denote the space of quadruples 
$(t_1,t_2,t_3,t_4)$ of $N\times N$ matrices.
The space $\overline\CV^{1,1}_N$ of monads is subset of $\CV_N$
satisfying \cad.
We write $\CV^{1,1}_N$ for the non-degenerated subset
of  $\overline\CV^{1,1}_N$

The remaining step is to identify 
the equivalence class of stable orbits
of the $GL(N,\msbm{C})$ action \idt\ (the GIT quotient).
We have a natural adjoint action $U(N)$ on
the quadruples $(t_1,t_2,t_3,t_4)$ preserving the obvious
norm. The 
momentum map (Hamiltonian) $\tilde\m_1$ 
is given by
\eqn\cae{
\tilde
\m_1(t_1,t_2,t_3,t_4) = [t_1,t_1^*] +[t_2,t_2^*] +[t_3,t_3^*]
+[t_4,t_4^*].
}
This form of the momentum map is follows from \idt.
Then we have the symplectic quotient, with the non-degeneracy
condition
\eqn\caeee{
{\CN}_N 
=\left({\CV}^{1,1}_N \cap \tilde\m^{-1}_1(0)\right)/U(N),
}
which is isomorphic to the GIT quotient,
$\CV^{1,1}_N //GL(N,\msbm{C})$.
We will call the above space the extended
moduli space of framed instantons. 

Finally, we note that the complex
equation \cad\ together with the real equation
\cae\ can be written in terms of hyperk\"{a}hler
momentum maps:
\eqn\hyperkk{
\eqalign{
\tilde\m_{\msbm{C}} &\equiv [t_1,t_2] +[t_3,t_4]=0,\cr
\tilde\m_{\msbm{R}} &\equiv [t_1,t_1^*] + [t_2,t_2^*] 
+ [t_3,t_3^*] + [t_4,t_4^*]=0.
}
}
After removing the non-degeneracy
condition we can write \caeee\ 
as
\eqn\caea{
\overline{\CN}_N = 
\left(\tilde\m_{\msbm{C}}^{-1}(0) 
\cap \tilde\m^{-1}_{\msbm{R}}(0)\right)/U(N).
}

\subsec{Dolbeault Equivariant Cohomology
and $S^1$ Actions}

Recall the natural adjoint $U(N)$ action on 
the quadruples $(t_1,t_2,t_3,t_4)$: 
\eqn\mmjh{
(t_1,t_2,t_3,t_4) \rightarrow (
gt_1 g^{-1},gt_2 g^{-1},g t_3 g^{-1}, gt_4 g^{-1}),
\qquad g \in U(N).
}The infinitesimal variation of this action
is given by
\eqn\tte{
\d t_i =[\d g ,t_i],\qquad
\d t_i^*=-[\d g,t_i^*]
}
for $i=1,2,3,4$.

In the space of the quadruples, we have an obvious
inner products and the K\"{a}hler potential $\eufm{k}$ 
is given by
\eqn\uuj{
\eufm{k} =-\Fr{1}{2}\tr(|t_1|^2 +|t_2|^2 +|t_3|^2 +|t_4|^2).
}
In addition to the adjoint  $U(N)$ actions, we
also have natural $S^1$-actions on the quadruples 
preserving the
natural norm.  
Such an action may be called
a hyper-Hamiltonian action
if it preserves the hyperk\"{a}hler
momentum maps \hyperkk. 
We have the following hyper-Hamiltonian $S^1\times S^1$ 
actions
\eqn\ttf{
(t_1,t_2,t_3,t_4) \rightarrow 
(e^{-i m^\pr\w^\pr}t_1,e^{+im^\pr\w^\pr}t_2,
e^{+im^\ppr\w^\ppr}t_3,
e^{-im^\ppr\w^\ppr}t_4),
}
where $\w^\pr$ and $\w^\ppr$ denote the generators
of $S^1_{m^\pr}$ actions on $(t_1,t_2,t_1^*,t_2^*)$
and of $S^1_{m^\ppr}$ action 
on $(t_3,t_4,t_3^*,t_4^*)$, respectively. 
Clearly the two $S^1$ actions
are disjoint at least for the equation \hyperkk.
Now the combined action of $U(N)\times S^{1}_{m^\pr}
\times S^{1}_{m^\ppr}$
can be summarized as
\eqn\tda{
\eqalign{
\d t_1=+im^\pr t_1 + [\d g,t_1],\cr
\d t_2=-im^\pr t_2 +[\d g,t_2],\cr
}\qquad
\eqalign{
\d t_3 = +im^\ppr t_3 +[\d g,t_3],\cr
\d t_4 = -im^\ppr t_4 +[\d g,t_4].\cr
}
}

It is convenient to formulate a Dolbeault version
of equivariant cohomology of the combined
action of $U(N)\times S^1_{m^\pr}\times S^1_{m^\ppr}$.
Using field theory language\foot{
This is the obvious cousin of the field theoretical
counterparts in \HPA.}, we define
the basic supersymmetry algebra generated by two
global supercharge $\bs$ and $\bbs$
\eqn\uaa{
\eqalign{
&\bs t_i = \p_i,\cr
&\bbs t_i=0,\cr
&\bs t_i^*=0,\cr
&\bbs t_i^*=\p^*_i,\cr
}\qquad
\eqalign{
&\bs\p_i =0,\cr
&\bbs\p_i=-i[t_0,t_i] 
- im^\pr t_i(\d_{1i}-\d_{2i})
- im^\ppr t_i(\d_{3i}-\d_{4i}),\cr
&\bs\p_i^*=-i[t_0,t_i^*] 
+ im^\pr t_i^*(\d_{1i}-\d_{2i})
+ im^\ppr t_i^*(\d_{3i}-\d_{4i}),\cr
&\bbs\p_i^*=0,\cr
}
}
where $i=1,2,3,4$,  $t_0=t_0^*$ is a $N\times N$ 
matrix, $m^\pr,m^\ppr$  are
real positive numbers and $\p_i$ and $\p_i^*$
are $N\times N$ matrix with anti-commuting
matrix elements. 
The algebra satisfies $\bs^2 =\bbs^2=0$
and
\eqn\uab{\eqalign{
&\{\bs,\bbs\}t_i =-i[t_0,t_i] - im^\pr t_i(\d_{1i}-\d_{2i})
- im^\ppr t_i(\d_{3i}-\d_{4i}),\cr
&\{\bs,\bbs\}t_i^* =-i[t_0,t_i^*] 
+ im^\pr t_i^*(\d_{1i}-\d_{2i})
+ im^\ppr t_i^*(\d_{3i}-\d_{4i}),
\cr
}
}
which show that the commutator $\{\bs,\bbs\}$
generate the infinitesimal $U(N)\times
S^1_{m^\pr}\times S^1_{m^\ppr}$ symmetry as given in \tda.
This is the equivariant extension of the
usual Dolbeault cohomology. 

We compute the equivariant extension of the
K\"{a}hler form;
\eqn\uuja{
-\Fr{i}{2}(\bs\bbs -\bbs\bs)\eufm{k}
=-\Fr{1}{2}
\tr\bigl( t_0\m_{\msbm{R}} 
+m^\pr(t_1t_1^*-t_2t_2^*)+m^\ppr(t_3t_3^*-t_4t_4^*)
-i\p_i\p_i^*\bigr),
}
where the repeated indices are summed over.
One can identify the first $3$ terms
with the momentum maps of $U(N)$, $S^1_{m^\pr}$ and
$S^1_{m^\ppr}$ actions, respectively.
The last term
$\Fr{i}{2}\tr\bigl( \p_i\p_i^*\bigr)$ can be identified
with the K\"{a}hler form.

There are other $S^1$ actions preserving either
$\tilde\m_{\msbm{C}}$ or $\tilde\m_{\msbm{R}}$
which will be considered later.

\subsec{$S^{1}_{m^\ppr}$ Action and Fixed Points}

The two sets of matrices
$(t_1,t_2)$ and $(t_3,t_4)$ are on equal
footing in our formulation.
The $4$ $N\times N$ matrices can be
viewed as $4$ endomorphisms
of $V=\msbm{C}^N$, i.e. $t_1,t_2,t_3,t_4
\in End(\msbm{C}^N)$, with $GL(N,\msbm{C})$ action.

To relate the ADHM constructions
and Sect.~$5.1$,
we will consider the $S^{1}_{m^\ppr}$ action
only and its fixed points. 
We set $m^\pr =0$ and $m^\ppr =m$.
The momentum map of $S^1_m$ action
is given by $\Fr{m}{2}\tr(t_3 t_3^* - t_4 t_4^*)$.
If we write an elemet of $U(N)$
as $e^{i t_0^a T^a}$, where $T^a$ denotes the
generator of Lie algebra of $U(N)$ in the
adjoint representation, we have
the following infinitesimal form of the combined action
of $U(N)\times S^1_m$
\eqn\tdaa{
\left\{
\eqalign{
\d t_1=i[t_0,t_1],\cr
\d t_2=i[t_0,t_2],\cr
\d t_1^*=i[t_0,t_1^*],\cr
\d t_2=i[t_0,t_2^*],\cr
}\right.
\qquad
\left\{
\eqalign{
\d t_3 = +im t_3 +i[t_0,t_3],\cr
\d t_4 = -im t_4 +i[t_0,t_4],\cr
\d t_3^* = -im t_3^* +i[t_0,t_3^*],\cr
\d t_4^* = +im t_4^* +i[t_0,t_4^*].
}\right.
}

Now we examine the various fixed point of
the $S^1$ action. The most obvious one
occurs when $t_3=t_4=0$. There are other
kinds of fixed points if and only if the $U(N)$
symmetry breaks down such that
the unbroken parts the symmetry
can undo the $S^1_m$ action. 
So the fixed points of $S^1_m$ action are identical
to the fixed point of the infinitesimal action
of $U(N)\times S^1_m$ described in \tdaa, together
with the symmetry breaking pattern.
Note that the above procedure
is equivalent to finding the fixed points of
the global supersymmetry $\bar\bs \p_i=\bs\p_i^*=0$.

For example the fixed point equation $\d t_3=0$
reduces to solving an eigenvalue problem
of operator $[t_0,\phantom{t_3}]$ acting
on $t_3$ with eigenvalue $-m$, $[t_0,t_3] = m t_3$.
Thus $t_0$ should be diagonalized, 
$t_0=U\phi U^\dagger$;
\eqn\raa{
\phi =diag(s_1,\ldots, s_N).
}   
We will order the diagonal components, using
the Weyl group, as $s_1\geq \ldots \geq s_N$.
Let $\{k_\ell\}$, $\ell =1,\ldots,n_p$
be a partition of $N$ by positive integers $k_\ell$,
$N=\sum_{\ell=1}^{n_p} k_\ell$.
To a given partitions, we associate
\eqn\rab{
\phi =bdiag(\a_1 I_{k_1},\ldots,\a_{n_p} I_{k_{n_p}}),
}
where $I_{k_\ell}$ denotes $k_\ell\times k_\ell$
unit matrix and $\a_1 > \ldots >\a_{n_p}$.
If, for example, $n_p$ is identical to $N$, we have
$k_\ell=1$ for all $\ell =1,\ldots,N$ and the
all eigenvalues $\a_\ell$ of $\phi$ in \raa\ are distinct.
The eignenvalues of $\phi$ determine the symmetry
breaking pattern. In other words we have
the eigenspace decomposition:
\eqn\eigens{
V= \oplus_{\ell}
V_{\ell},\qquad\hbox{with}\quad
N=\sum_{\ell=1}^{n_p} k_\ell,\quad
V_{\ell} = \msbm{C}^{k_\ell}.
}

For a fixed partition $\{k_\ell\}$ with $\ell =1,\ldots,n_p$,
the fixed points of the $S^1$ action 
depending on \rab\ are easily
determined.
The second set of equations in \tdaa\ implies
$t_3$ becomes block upper trianglular, provided that 
$m > 0$. Then  $t_3$ has $n_p\times n_p$
blocks and $(\ell,\ell^\pr)$-th 
block denoted by  $\s_{\ell,\ell^\pr}$
is a $k_\ell \times k_{\ell^\pr}$ matrix
with $\s_{\ell,\ell^\pr} = 0$ for $\ell \geq \ell^\pr$.
It is also easy to see that not all of the 
$\s_{\ell,\ell^\pr}$ with $\ell > \ell^\pr$
are non-vanishing.
In general a $k_\ell\times k_{\ell^\pr}$
matrix $\s_{\ell,\ell^\pr}$ vanishes if 
\eqn\abac{
\a_\ell -\a_{\ell^\pr} \neq m.
}
Otherwise there is no restriction.
Note that we have arranged that $\a_\ell > \a_{\ell^\pr}$
if $\ell > \ell^\pr$ and $m$ is a positive number. 
Similarly, 
$t_4$ has $n_p\times n_p$ blocks
and the $(\ell,\ell^\pr)$-th 
block denoted by  $\pi_{\ell,\ell^\pr}$
is a $k_\ell \times k_{\ell^\pr}$ matrix
with $\pi_{\ell,\ell^\pr} = 0$ for $\ell \leq \ell^\pr$.
When the condition
$\a_\ell -\a_{\ell^\pr} =m$ is satisfied, the matrix $\pi_{\ell^\pr,\ell}$ has
no restriction. On the other hand,
if $\a_\ell -\a_{\ell^\pr} \neq m$ the matrix
$\pi_{\ell^\pr,\ell}$
is identically zero.
The first set of equations in \tdaa\ implies
$t_1$ and $t_2$ become block diagonal
with $n_p$ blocks.
We will denote the block
elements of $t_1$ and $t_2$
by $z_\ell$ and $w_\ell$, repectively; 
both are $k_\ell\times k_\ell$ matrices.

The above discussions can be summarized as
follows. Under the eigenspace decompositions
\eigens, the original endomorphism of $V$
is decomposed into $End(V_\ell)$ and 
$Hom(V_\ell, V_{\ell^\pr})$
with $\ell,\ell^\pr \in
1,\ldots, n_p$.
Then we have identifications: 
$z_\ell,w_\ell \in End(V_\ell)$,
$\s_{\ell,\ell^\pr} \in Hom(V_\ell, V_{\ell^\pr})$
and $\pi_{\ell^\pr,\ell} \in Hom(V_{\ell^\pr}, V_{\ell})$
with $\ell < \ell^\pr$.

\subsubsec{Examples}

Now we consider some examples of fixed points.

\lin{Example (1)}

Consider all eigenvalues in \raa\ are the same.
The only solution of \tdaa\ is $t_3=t_4=0$.
There are no restrictions to $t_1$ and $t_2$.
The equation \hyperkk\ reduces to
\eqn\uha{
[t_1,t_2]=0,\qquad [t_1,t_1^*] + [t_2,t_2^*]=0.
}
The original $U(N)$ symmetry is unbroken.
One obvious family of solutions of above
for $t_1, t_2$ are diagonalized
matrices; 
$$
t_1 = diag(z^1,\ldots,z^N),\quad
t_2 = diag(w^1,\ldots,w^N)
$$
One can regard the eigenvalues as the positions
of $N$ unordered points in $\msbm{C}^2$.
So the fixed point locus is $S^{N}(\msbm{C}^2)$.
Note that $\Fr{1}{N}(\tr t_1,\tr t_2)$ can
be interpreted as the center of mass coordinates
in $\msbm{C}^2$. This is the ADHM description for
$N$ point-like instantons.

\lin{Example (2)}

We consider a partition
$N = k +n$ and $\phi = bdiag(\a I_k, (\a -m) I_n)$
leading to the following eigen space decomposition;
\eqn\fafsdf{
V = V_1 \oplus V_2,\qquad V_1 =\msbm{C}^k,\quad
V_2 =\msbm{C}^n,
}
The original $U(N)$ ($GL(N,\msbm{C})$)
symmetry is broken to $U(k)\times U(n)$
($GL(k,\msbm{C})\times GL(n,\msbm{C})$).
We have
\eqn\waste{
t_1=\left(\matrix{z_1&0\cr 0&z_2}\right),
\qquad
t_2=\left(\matrix{w_1&0\cr 0&w_2}\right)
,\qquad
t_3 =\left(\matrix{0&\s_{1,2}\cr 0&0}\right)
,\qquad
t_4 =\left(\matrix{0&0\cr \pi_{2,1}&0}\right),
}
where 
$\s_{1,2}$ is $k\times n$ and $\pi_{2,1}$ 
is $n\times k$ matrices.
In other words, we have
$z_1,w_1 \in End(V_1)$, $z_2,w_2 \in End (V_2)$,
$\s_{1,2}\in Hom (V_2, V_1)$ and $\pi_{2,1}
\in Hom(V_1,V_2)$.
We have
\eqn\xxa{\left\{\eqalign{
&[z_1, w_1] + \s\pi=0,\cr
&[z_1,z_1^*] +[w_1,w_1^*] + \s\s^* 
-\pi^*\pi=0,
}\right.\quad
\left\{\eqalign{
&[z_2, w_2] -\pi\s=0,\cr
&[z_2,z_2^*] +[w_2,w_2^*] 
 +\pi\pi^*-\s^*\s=0,\cr
}\right.
}
Note that the first and the second sets of equations describe
$k$ $SU(n)$ and $n$ $SU(k)$ instantons, repectively.
Assumming the non-degeneracy condition,
the fixed point locus is
$
\left(\left(\CM_{n,k}\right)\times \left(\CM_{k,n}\right)
\right)/S_2,
$
where $S_2$ for $k\neq 0$, 
is the remnant the Weyl group of $U(N)$.

\bigskip
     \epsfxsize=5truecm
\centerline{
\epsfbox{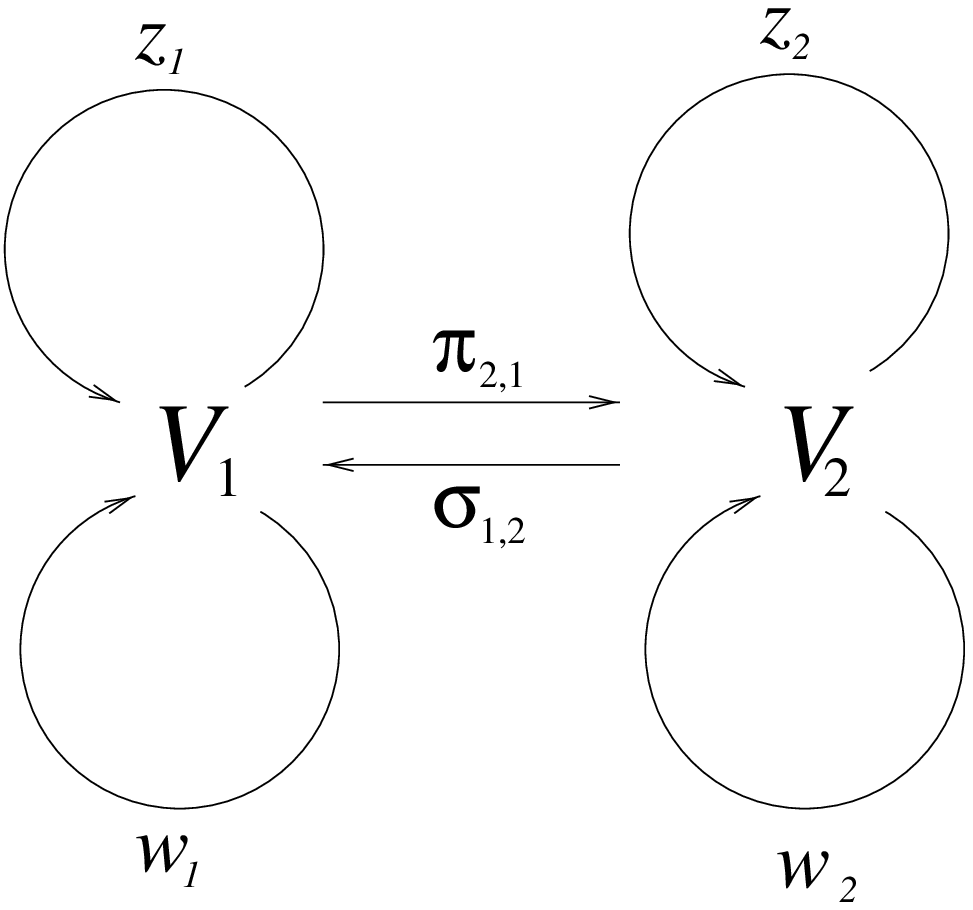}
}
\medskip
{\footskip14pt plus 1pt minus 1pt \footnotefont
{\bf Fig.~2.} 
The example (2):
$V= V_1 \oplus V_2$ where  
$V_1=\msbm{C}^k$, $V_2=\msbm{C}^n$ with $N=n+k$, 
$z_1,w_1 \in End(V_1)$,
$z_2,w_2 \in End(V_2)$,
$\s_{1,2}\in Hom(V_2,V_1)$ and $\pi_{2,1}\in Hom(V_1,V_2)$.
This is to be compared with the usual ADHM description
in Fig.~1. In a latter section we will interpret
this diagram in terms of $N$ D-instatons
confined on two $4$-dimensional hyperplanes in $10$-dimensions.
The endomorphisms and homomorphisms will be
interpreted as open strings joining the various D-instantons. 
}
 \bigskip

Removing the non-degeneracy condition the
first equation in \xxa\ has lower strata
$\CM_{N-k,k-\ell}\times S^\ell(\msbm{C}^2)$
for $\ell = 1,\ldots k$. That is $\s$
and $\pi$ degenerate to $(k-\ell)\times (N-k)$
and $(N-k)\times (k-\ell)$ matrices, respectively. 
Then the second
equation describe $N-k$ $SU(k-\ell)$ non-degenerated
instantons. Then we have
$
\left(
\left(\CM_{N-k,k-\ell}\times S^\ell(\msbm{C}^2)\right)
\times\left( \CM_{k-\ell,N-k}\right)\right)/S_2$.
Now the $N-k$ $SU(k-\ell)$ instantons can also
degenerate to $N-k -\ell^\pr$ 
$SU(k-\ell)$ instantons,
for $\ell=1,\ldots, N-k$.
Then the first equation describes
$k-\ell$ $SU(N-k-\ell^\pr)$ instantons;
\eqn\stra{
\left(\left(\CM_{N-k-\ell^\pr,k-\ell}\times 
S^\ell(\msbm{C}^2)\right)
\times \left(\CM_{k-\ell,N-k-\ell^\pr}
\times S^{\ell^\pr}(\msbm{C}^2)\right)
\right)/S_2
}
Consequently the fixed point locus without the
non-degeneracy condition is
\eqn\strb{
\bigcup_{\matrix{{}_{\ell=0,\ldots,k} \cr 
{}_{\ell^\pr=0,\ldots,N-k}}}
\biggl(
\biggl(\CM_{N-k-\ell^\pr,k-\ell}\times S^\ell(\msbm{C}^2)
\biggr)
\times \biggl(\CM_{k-\ell,N-k-\ell^\pr}
\times S^{\ell^\pr}(\msbm{C}^2)\biggr)
\biggr)/S_2.
}

Here we obtain two sets of Yang-Mills instantons
correlated with each other in somewhat misterious
way. In the later sections, we will intepret
our extended monads to describe D-instantons.
According to D-instantons picture, the above
system describes $k$ and $n$ D-instantons
confined in two parallel $4$-dimensional hyperplanes
in ten-dimensional space.
The strange correlation will be explained by
strings. 
For an observer in a hyperplane, the above
system is viewed as
\eqn\strac{
\bigcup_{\matrix{{}_{\ell=0,\ldots,k} \cr 
{}_{\ell^\pr=0,\ldots,N-k}}}
\biggl(\CM_{N-k-\ell^\pr,k-\ell}\times S^\ell(\msbm{C}^2)
\biggr)
= \bigcup_{\ell^\pr =0}^{N-k}\overline
\CM_{N-k-\ell^\pr, k-\ell}. 
}
We may interpret the above moduli space
as a double stratification of the moduli space
of $k$ $SU(N-k)$ instantons, one
way by $\ell$ number of point-like instantons
and the other way by $\ell^\pr$ number
of flat-factors of $SU(N-k)$ bundles.\foot{
A flat factor is the reducible connection 
$E = E_0\oplus L_f$, where $L_f$ has flat
$U(1)$ connections. 
}
Then the system described by \strb\
is manifestly reciprocal. The $S_2$ action
exchanges two $\msbm{C}^2$ and the rank of gauge
group and the instanton numbers as well as
the two stratifications defined by the
flat factors and point-like instantons.
Analogous things happen in case of
Fourier-Nahm transformation of instantons
on a flat $4$-torus \BB\DK.

\lin{Example (3)}
 
Let $\phi$ is given by \rab. 
The maximum non-vanishing number
of $\s_{\ell,\ell^\pr}$ is $n_p -1$
when the values $\a_\ell$ are evenly spaced
by $m$, i.e., $\a_1,\a_1 -m,\a_1 -2m,\ldots, 
\a_1 -(n_p -1)m$.
Then the non-vanishing
elements of $t_3$ are $\s_{\ell,\ell+1}$, 
$\ell =1,\ldots,n_p$. Similarly, 
the non-vanishing elements of $t_4$
are $\pi_{\ell,\ell -1}$.
If $n_p=3$, for example, we have 
\eqn\whym{
t_1 = \left(\matrix{
z_1&0&0\cr
0&z_2&0\cr
0&0&z_3
}\right),
\qquad
t_3 =\left(\matrix{
0&\s_{1,2}&0 \cr 
0&0&\s_{2,3}\cr 
0&0&0 
}\right),
\qquad
t_4 =\left(\matrix{
0&0&0 \cr 
\pi_{2,1}&0&0\cr 
0&\pi_{3,2}&0}
\right),
}
with the following diagram;
\bigskip
     \epsfxsize=8truecm
\centerline{
\epsfbox{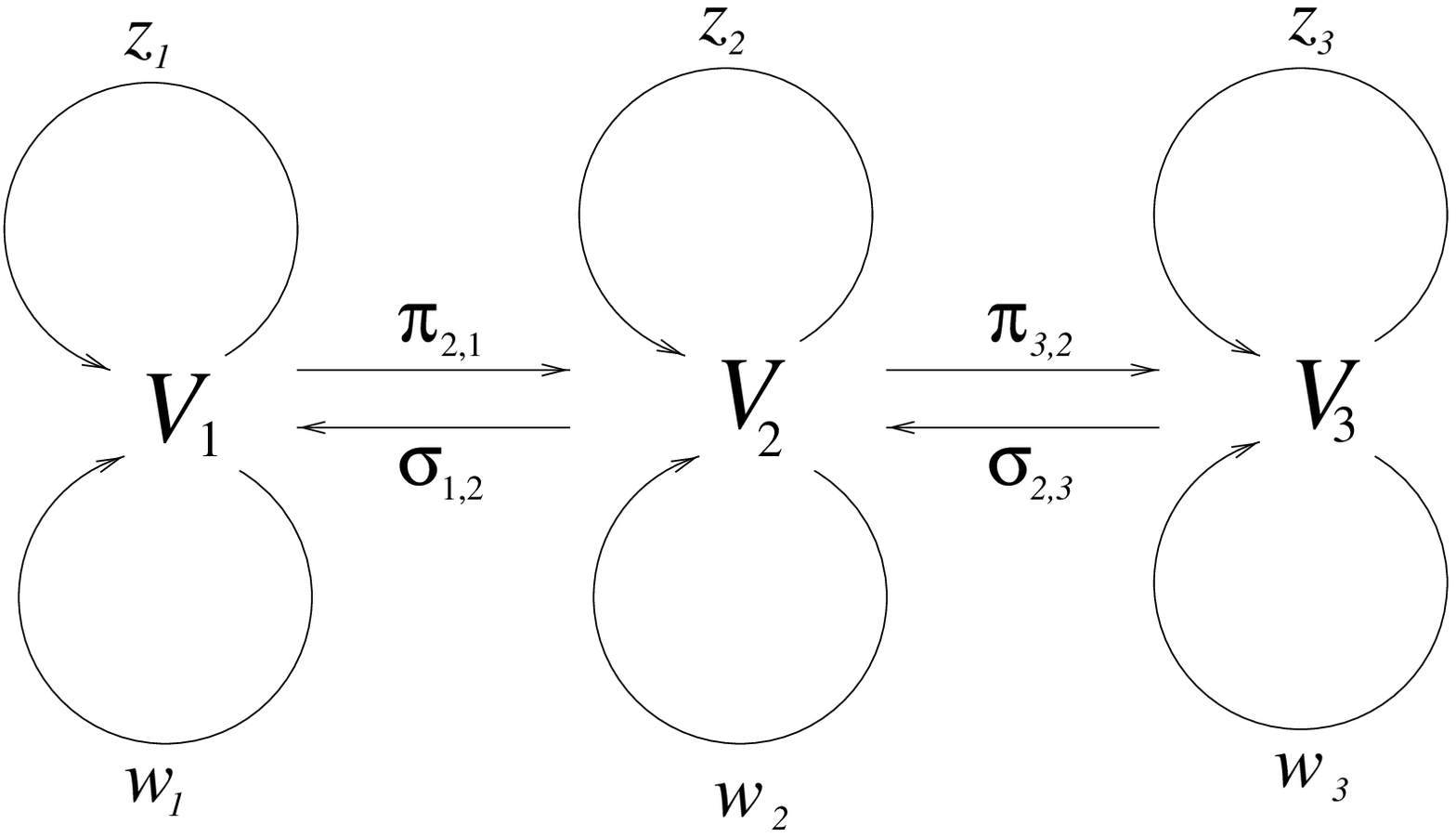}
}
\medskip
{\footskip14pt plus 1pt minus 1pt \footnotefont
{\bf Fig.~3.} The eigen space decomposition
$V = V_1 \oplus V_2 \oplus V_3$, 
$V_\ell = \msbm{C}^{k_\ell}$ with $\sum_{\ell =1}^3 k_\ell= N$.
}
 \bigskip

The equations
\hyperkk\ becomes
\eqn\exo{\eqalign{
[z_1, w_1] + \s_{1,2}\pi_{2,1}=0,\cr
[z_1,z_1^*] +[w_1,w_1^*] + \s_{1,2}\s_{1,2}^* 
-\pi_{2,1}^*\pi_{2,1}=0,\cr
.....,\cr
[z_\ell, w_\ell] + \s_{\ell,\ell+1}\pi_{\ell+1,\ell} 
-\pi_{\ell,\ell-1}\s_{\ell-1,\ell}=0,\cr
[z_\ell,z_\ell^*] +[w_\ell,w_\ell^*] +
\s_{\ell,\ell+1}\s_{\ell,\ell+1}^* 
-\pi_{\ell+1,\ell}^*\pi_{\ell+1,\ell} 
-\s_{\ell-1,\ell}^*\s_{\ell-1,\ell} 
+\pi_{\ell,\ell-1}\pi_{\ell,\ell-1}^*=0,\cr
.....,\cr
[z_{n_p}, w_{n_p}] - \pi_{n_p,n_p-1}\s_{n_p-1,p}=0,\cr
[z_{n_p},z_{n_p}^*] +[w_{n_p},w_{n_p}^*] 
+ \s_{n_p-1,n_p}^*\s_{n_p-1,n_p} 
-\pi_{n_p,n_p-1}\pi_{n_p-1,n_p}^*=0.
}
}
The first two equations of \exo\ are 
precisely the ADHM equations for $SU(k_2)$ instantons
with instanton number $k_1$ and last two equations
describe $SU(k_{n_p})$ instantons with
instanton numbers $k_{n_p-1}$.
The two equations in the middle of \exo\
may be identified with the ADHM equations
for the instanton number $k_\ell$ with gauge group
$SU(k_{\ell+1})\times SU(k_{\ell-1})$.
Now the fixed point locus is much more complicated.
One can do the similar analysis as the example (2).
We only note that the above system has $S_{n_p}$
symmetry.

We see that the series of ADHM equations in \exo\
and the corresponding diagram, for example Fig.~3.,
have certain similarity with ADHM description
of instantons on ALE space and the quiver diagram
\KN\Nakajima\DOM.

\lin{Example (4)}

Let $\phi$ is given by \rab. When  
$|\a_{\ell} -\a_{\ell^\pr}|\neq m$ for all $\ell,\ell^\pr$, 
all the matrices $\s_{\ell,\ell^\pr}$ and
$\pi_{\ell,\ell^\pr}$ vanish, i.e.~$t_3 = t_4=0$. 
Then the equations \hyperkk\ becomes
\eqn\eba{
\eqalign{
[z_\ell,w_\ell] =0, \cr
[z_\ell,z_\ell^*] +[w_\ell,w_\ell^*]=0,
}
}
where $z_\ell$ and $w_\ell$ are $k_\ell\times k_\ell$
matrices. This is the completely degenerated
limit of example (3).
The system described by \eba\ corresponds to
$n_p$ paralell $4$-dimensional hyperplanes
containing $k_\ell$ point-like instantons
in $\ell$-th hyperplane.

\subsec{$S^1_{m_0}$ Action on ADHM Data and 
the Stratification}

In the previous subsection, we recovered
the usual ADHM construction by considering 
fixed points of the natural $S^1_m$ action.
Within a fixed points locus, we still have
the $S^1$ action. So we will return to
the case reviewed in Sect.~$2.2$ and use
the same notation.
We had the quadruples $(\t_1,\t_2,a,b)$,
with $GL(k,\msbm{C})$ action
\eqn\nglk{
(\t_1,\t_2,a,b)\rightarrow 
(p\t_1 p^{-1},p\t_2 p^{-1},a p^{-1},pb),
\qquad p\in GL(k,\msbm{C}).
} 
We had momentum maps;
\eqn\realm{
\eqalign{
&\m_{\msbm{C}}=[\t_1,\t_2] + ba,\cr
&\m_{\msbm{R}} 
  = [\t_1,\t_1^*] +[\t_2,\t_2^*] +b b^* - a^*a,
}
}
and the quotient
$(\m^{-1}_{\msbm{R}}(0)\cap 
\m^{-1}_{\msbm{C}}(0)/U(k))=\overline{\hat \CM}_{k,n}$.
We consider the following $S^1_{m_0}$ action
on the above quotient space;
\eqn\nga{
(\t_1,\t_2, a, b)\rightarrow (\t_1,\t_2, e^{i m_0\w_0}a,
e^{-im_0\w_0}b),
}
This is analogous to the $S^1$ action considered
in Sect.~$3.2$ and preserves both
real and complex momentum maps. 

Let $h$ denote
a infinitesimal generator of the $U(k)$
acting as \nglk. The combined action of
$U(k)$ and $S^1$ is given by
\eqn\ngb{
\eqalign{
\d \t_1 = i[h,\t_1],\cr
\d \t_2 = i[h,\t_2],\cr
}
\qquad
\eqalign{
\d a = +im a - iah,\cr
\d b = -im b + ihb.\cr
}
}
The above transformation law should be
compared with \tdaa.
One now repeat what we did in Sect.~$3.1.2$.
The momentum map of the $S^1_{m_0}$ is given by 
$\Fr{m_0}{2}\tr(a^*a - bb^*)$.

Let $\{k_\ell\}$, $\ell =1,\ldots,n_p$
be a partition of $k$, 
$k=\sum_{\ell=1}^{n_p} k_\ell$.
For a given partition, we associate
a diagonalization of $k\times k$ matrix $h$;
\eqn\rab{
h =bdiag(\a_1 I_{k_1},\ldots,\a_{n_p} I_{k_{n_p}}),
}
with all $\a_\ell$ are distinct.
{}From $\d \t_1=\d \t_2 =0$, $\t_1$ and $\t_2$
becomes $n_p\times n_p$ block diagonal matrices;
$\t_1 = bdiag(z_1,z_2,\ldots,z_{n_p})$,
$\t_2 = bdiag(w_1,w_2,\ldots,w_{n_p})$.
For generic choice of $h$, $a$ and $b$ are zero
to solve $\d a =\d b =0$.

Consider the
case $k = (k-l) + l$, the condition $\d a =0$
is
\eqn\boring{ 
m\left(\matrix{a_{1,1} & a_{1,2}}\right)
- \left(\matrix{a_{1,1} & a_{1,2}}\right)
\left(\matrix{\a_1 & 0\cr 0 &\a_2}\right)  
= \left(\matrix{
(m-\a_1)a_{1,1} & (m-\a_2)a_{1,2}}
\right)=0,
}
where $a_{1,1}$ and $a_{1,2}$ are $n\times (k-l)$
and $n\times l$ matrices.
Non-trivial solutions occur for either $\a_1 =m$
or $\a_2 =m$. We can choose $\a_1 =m$ up to
Weyl group of $U(k)$.
Then, the condition $\d a=\d b =0$ are solved
by
\eqn\boringb{
a = \left(\matrix{a_{1,1} &0}\right),
\qquad
b = \left(\matrix{b_{1,1}\cr 0}\right).
}
Now we have
\eqn\boringc{
\m_{\msbm{C}} = 0
\rightarrow
\left(\matrix{
[z_1,w_1] +b_{1,1}a_{1,1} & 0\cr
0 & [z_2,w_2]}\right)=0.
}
So the fixed point locus is
$\hat \CM_{k-l,n}\times S^l(\msbm{C}^2)$.
If we consider the case  $k = \sum_{\ell=1}^{n_p}k_\ell$,
we will get one
of stratum of $\overline{\hat\CM}_{k,n}$ as  the fixed
point locus.

\newsec{The Relations with ${\cal N}=4$ Super-Yang-Mills Theory}

Now we will explain the original motivation
of this paper. 
To begin with, we rewrite the
equations we considered in the previous sections;
\eqn\summ{
\eqalign{
\tilde\m_{\msbm{C}}=[t_1,t_2] +[t_3,t_4]=0,\cr
\tilde\m_{\msbm{R}}=
[t_1,t_1^*] + [t_2,t_2^*] + [t_3,t_3^*] +[t_4,t_4^*] =0,\cr
}
}
and
\eqn\summb{\eqalign{
[t_5, t_1] =0,\cr
[t_5, t_2] =0,\cr
}\qquad
\eqalign{
m t_3 + [t_5,t_3]=0,\cr
m t_4 - [t_5,t_4]=0,\cr
}
\qquad
\eqalign{
[t_5,t_5^*]=0,
}
}
where we replaced $t_0$, which
generate the $U(N)$ symmetry,
with $t_5$, which generates 
the $GL(N,\msbm{C})$.\foot{Here 
$t_5$ and $t_5^*$ are 
They originated from the complex Higgs scalar in
the ${\cal N}=2$ super-Yang-Mills theory, usually
denoted by $\phi$ and $\bar\phi$. 
In the twisted
theory it is more natural to regard them as
two independent real scalars.
Note that $\tilde\m_{\msbm{R}}$
is the $U(N)$ moment map.
}
We also supplement  the two sets of equations \summ\
and \summb\ with
\eqn\mmde{
[t_2,t_3]+ [t_1,t_4]=0,
\qquad
[t_3,t_1]+ [t_2,t_4]=0.
}
These additional equations also do not
alter our previous discussions.
For example consider a branch
of the fixed points given by a partition
$N= k + n$ such that
\eqn\uiop{
t_1 =\left(\matrix{z_1 & 0\cr 0 & z_2}\right),
\qquad
t_3 = \left(\matrix{0 & \s\cr 0 & 0}\right),
\qquad
t_4 = \left(\matrix{0 & 0\cr \pi & 0}\right)
}
Then \mmde\ reduces
to
\eqn\ujop{
\eqalign{
z_2 \s  = \s z_1, \cr
z_1\pi = \pi z_2,\cr
}\qquad
\eqalign{
z_1\s = \s z_2,\cr
z_2\pi = \pi z_1,\cr
}
}
For example, the equation for $\s$ implies
that $\s$ should be invariant under
the combination of the 
natural left action of $GL(k,\msbm{C})$ and the right
action $GL(n,\msbm{C})$. 
This is entirely consistent with our previous
discussions since we reduced everything
to the fixed points of the group actions.
For this new but equivalent set of equations
\summ, \summb\ and \mmde, the higher
dimensional origin of our construction
becomes obvious.

\subsec{Dimensional reduction of Vafa-Witten equations}

Now we consider a twisted ${\cal N}=4$ super-Yang-Mills
theory studied by Vafa and Witten \VW. 
In the paper \DPS, we systematically
studied the ${\cal N}=4$ theory on a K\"{a}hler surface 
and examined how the topological theory changes
as one breaks the supersymmetry down to ${\cal N}=2$
and ${\cal N} =1$. Here I will summarize some relevant
results of \VW\DPS.

On a K\"{a}hler surface, the model can be
reformulated such that there are $4$ topological
charges. The resulting theory is an example
of the balanced topological field theory \DM\
with the Dolbeault version of the $\CG$-equivariant
cohomology. In this model one can introduce
${\cal N}=2$ preserving bare mass to the adjoint 
hyper-multiplet. It turns out that introducing
the bare mass is equivalent to combining 
natural $S^1$ actions on the two (twisted)
complex scalars in the hyper-multiplet
with the $\CG$-equivariant cohomology.
Consider the theory with $U(N)$ gauge
group. The equations of the minimum of the bosonic
part of the action ), after turning
on the bare mass $m$ (equivalently, after
breaking ${\cal N}=4$ down to ${\cal N}=2$ supersymmetry) are given
by
\eqn\daa{\eqalign{
F^{0,2}_{A_0} + [t_3, t_4] =0,\cr
F_{A_0}\wedge\o +[t_3,t_3^*] +[t_4, t_4^*]\o^2 =0,\cr
}
\qquad
\Dp t_5 =\Dpp t_5= 0,
\qquad
\Dpp^* t_3 + \Dpp t_4=0,
}
together with 
\eqn\daato{
\eqalign{
m t_3 + [t_5,t_3]=0,\cr
m t_4 - [t_5,t_4]=0,\cr
}
\qquad
\eqalign{
[t_5,t_5^*]=0.
}
}
Note that the above two sets of equations \daa\ and
\daato\ are fixed points of the global supersymmetry
of the twisted ${\cal N}=4$ super-Yang-Mills
after turning on the bare mass.

Here we write $F_{A_0}$ for the trace free part,
i.e.~$F_A = F_{A_0} +\Fr{1}{N}\tr F_A\otimes I_N$.
We also have $\tr F_A^+ = \o^+$ for a self-dual
two form $2\pi i\o^+=c_1(E)$.
One may identify $t_5$
with a scalar in the ${\cal N}=2$ vector
multiplets, $t_3$ and $t_4$ as the two complex
scalars in the ${\cal N}=2$ hyper-multiplet.
After twisting $t_3$ becomes the holomorphic two
form. Note that every scalar fields carry the adjoint
representation of $U(N)$. The equation \daa\ is
nothing but the Vafa-Witten equation in the complex
coordinates.
The equations \daa\ are invariant under
the $S^1$ actions 
$(t_3,t_4)\rightarrow (e^{i m\w}t_4,e^{-i m\w}t_4)$. 
This originates from the unbroken global
symmetry of the twisted ${\cal N}=4$ theory \VW\
and identical to the $S^1_m$ action studied
in Sect.~$3.1$.

Together with $4$ components of the connection,
we have $10$ matrix-valued forms.
If we do dimensional reduction of the above equations
down to $0$-dimension,
we precisely get our  equations \summ\summb\mmde\
for extended monads.
In other words, the dimensional reduction
of ${\cal N}=2$ preserving ${\cal N}=4$ super-Yang-Mills
theory reduces the infinite dimensional 
$\CG\times S^1$-equivariant
cohomology of the space of connections
with additional matter fields 
to the finite dimensional $U(N)\times S^1$-equivariant
cohomology of the extended monads.
And the both equation describe Yang-Mills instantons.\foot{
Note that the Vafa-Witten equation \daa\ on
$\msbm{R}^4$ enjoys certain vanshing theorem 
such that it reduces to the ASD equation \VW.}
This is an amazing property stems 
from  reciprocity.

\subsubsec{
Perturbation to ${\cal N} =1$ theory
and the resolution of the singularities}

In the paper of Vafa and Witten \VW, a perturbation
of ${\cal N}=4$ theory to ${\cal N} =1$ theory was crucial.
If we break the supersymmetry further down 
to ${\cal N} =1$, the classical vacua are classified
by the complex conjugacy classes of homomorphisms
of the $SU(2)$ Lie algebra to that of $U(N)$ \VW. 
In that case, the first column of \daa\
is changed in a very interesting way;
\eqn\pertu{\eqalign{
F^{0,2}_{A_0} 
+ [t_3,t_4] + t_5^0\o^{0,2} = 0,\cr
\tr F^{0,2}_A + \tr t_5\o^{0,2} = (\o^+)^{0,2},
}}
where $\o^{0,2}$ is the holomorphic two-form
which corresponds to the ${\cal N} =1$ preserving
bare mass and $t_5^0$ is the trace-free
part of $t_5$. In the paper \DPS, we use
the above to recover Witten's judicious perturbation of
the Abelian Seiberg-Witten equation  \WittenB.
In other words, the perturbation considered by
Vafa and Witten \VW\ becomes identical to
Witten's perturbation of abelian Seiberg-Witten
equation at a fixed point of the $S^1$ action.
Such perturbation 
leads to a factorization
of Seiberg-Witten class which is crucial
in finding all the basic classes \WittenB. Mathematically
such perturbation ensure the moduli space
of abelian Seiberg-Witten monopoles behaves well.

Now we dimensionally reduce the perturbed
equation down to zero.  Then, the equations 
in \summ\
becomes
\eqn\judicious{
\eqalign{
\tilde\m_{\msbm{C}} &= [t_1,t_2] + [t_3,t_4] = 
-t^0_5,\cr
\tilde\m_{\msbm{R}} &=
[t_1,t_1^*] + [t_2,t_2^*] + [t_3,t_3^*] +[t_4,t_4^*] =0,\cr
}
}
Note that the real equation in \summ\ 
and the remaining
equations \summb\mmde\ remain unchanged.

Now we consider solutions of
\judicious\ together with equations  \summb\mmde.
This is equivalent to considering the
fixed points of
$S^1_m$ action discussed in Sect.~$3.2$.
For simplicity, we consider the
case of example (2), we have (the modification
for \xxa);
\eqn\xxap{\left\{\eqalign{
&[z_1, w_1] + \s\pi=-\a I_k,\cr
&[z_1,z_1^*] +[w_1,w_1^*] + \s\s^* 
-\pi^*\pi=0,
}\right.\quad
\left\{\eqalign{
&[z_2, w_2] -\pi\s=\Fr{k\a}{n} I_n,\cr
&[z_2,z_2^*] +[w_2,w_2^*] 
 +\pi\pi^*-\s^*\s=0,\cr
}\right.
}
with $(1+k/n)\a= m$.
For example, the first equation
says the fixed point locus
is
\eqn\resol{
\overline\CM^{\e_{\msbm{C}}\a}_{n,k}=
\m_{\msbm{R}}^{-1}(0)\cap \m_{\msbm{C}}^{-1}(-\a 
I_k)
/U(k)\equiv 
\m_{\msbm{C}}^{-1}(-\a I_k)
//GL(k,\msbm{C}).
} 
This fits nicely with the variation of GIT quotients
such that the above can be viewed as 
a resolution of the singularities \Kirwan\ of 
the completed moduli space of framed instantons:
\eqn\orign{
\overline\CM_{n,k}=
\m_{\msbm{R}}^{-1}(0)\cap \m_{\msbm{C}}^{-1}(0)
/U(k) = \bigcup_{\ell=0}^{k}\left(\hat \CM_{k-\ell,N-k}\times
S^\ell(\msbm{C}^2)\right).
}
It is amusing to see that the same prescription
in ${\cal N}=4$ SYM theory leads both the moduli space
of Seiberg-Witten monopoles and the
completed moduli space of framed instantons.

\subsubsec{$S^1$ Actions invisible 
in the Vafa-Witten equation}

We saw that the extended monads considered in Sect.~$3.1$
are directly related with the ${\cal N}=4$ super-Yang-Mills
theory in $4$-dimensions. The $S^1_m$ action consider
in Sect.~$3.2$ was precisely motivated by the ${\cal N}=4$
SYM theory, which has the same global symmetry
and the bare mass is related to the generator
of the same $S^1_m$ action.
A crucial difference from the Vafa-Witten
equation \daa\ and the equations for the extended
monads \summ\ with \judicious\ is that the later has
another $S^1\times S^1$ action;
\eqn\asone{
(t_1,t_2,t_3,t_4) \rightarrow 
(e^{i\vr^\pr}t_1,e^{i\vr^\ppr}t_2,
e^{i\vr^\pr}t_3,e^{i\vr^\ppr}t_4).
}
This action preserve $\tilde\m_{\msbm{R}}$
while the complex momentum map gets the
phase factor $\tilde\m_{\msbm{C}} \rightarrow
e^{i(\vr^\pr +\vr^\ppr)}\tilde\m_{\msbm{C}}$.\foot{ 
The choice of 
the $S^1_{\vr^\pr} \times S^1_{\vr^\ppr}$ action has made to
preserve the equations \mmde, which is
the dimensional reduction of the Dirac equations
for the two complex fields in the hyper-multiplet.
The later can be also viewed as a momentum map.}

Now the space of the extended monads 
has $U(N)\times S^1_m\times S^1_{\vr^\pr} 
\times S^1_{\vr^\ppr}$, which preserve
the K\"{a}hler potential $\eufm{k}$ given
by \uuj. This leads to the following
modification of the 
basic supersymmetry algebra \uaa
\eqn\uaab{
\eqalign{
&\bs t_j = \p_j,\cr
&\bbs t_j=0,\cr
&\bs t_j^*=0,\cr
&\bbs t_j^*=\p^*_j,\cr
}\qquad
\eqalign{
&\bs\p_j =0,\cr
&\bbs\p_j=-i[t_0,t_j] 
+ i\vr^\pr t_j(\d_{1j}+\d_{3j})
+ i\vr^\ppr t_j(\d_{2j}+\d_{4j})
- im t_j(\d_{3j}-\d_{4j}),\cr
&\bs\p_j^*=-i[t_0,t_j^*] 
- i\vr^\pr t_j^*(\d_{1j}+\d_{3j})
- i\vr^\ppr t_j^*(\d_{2j}+\d_{4j})
+ im t_j^*(\d_{3j}-\d_{4j}),\cr
&\bbs\p_j^*=0,\cr
}
}
where $j=1,2,3,4$.
Then one find the equivariant K\"{a}hler form 
as
\eqn\uujaa{
-\Fr{i}{2}(\bs\bbs -\bbs\bs)\eufm{k}
=-
\Fr{1}{2}
\tr\bigl( t_0\tilde\m_{\msbm{R}} 
+m(t_3t_3^*-t_4t_4^*)
+\vr^\pr(t_1t_1^*+t_3t_3^*)
+\vr^\ppr(t_2t_2^*+t_4t_4^*)
-i\sum_j \p_j\p_j^*\bigr).
}
One can identify the first $4$ terms
with the momentum maps of $U(N)$, $S^1_m$,
$S^1_{\vr^\pr}$ and
$S^1_{\vr^\ppr}$ actions, respectively. 
Note that the Hamiltonian,
\eqn\hamill{
-\Fr{1}{2}
\tr\bigl(
\vr^\pr(t_1t_1^*+t_3t_3^*)
+\vr^\ppr(t_2t_2^*+t_4t_4^*)\bigr), 
} 
of $S^1_{\vr^\pr}\times S^1_{\vr^\ppr}$
action is proportional to the K\"ahler
potential $\eufm{k}$ for $\vr^\pr = \vr^\ppr$. 
We can use the above as a Morse function
to study homology. The critical
points of such a Morse function
are identical to the fixed point of $\bs$
and $\bbs$ in \uaab. 
Physically, including those $S^1$ actions 
corresponds to giving masses to all the
fields. This is not possible in general for its
field theoretical cousins without breaking
the gauge symmetry.
Now we can conclude that any homological
problem concerning our extended moduli space
can be localized to the problem on the fixed
point locus of above.
This additional symmetry should greatly simplify
any approach based on the space of monads.

\subsec{A  Matrix Model}

Physically, we can use \uujaa\ 
as an action functional.
Then we have a supersymmetric matrix integral
analog to the path integral integral. 
Provided that
we restrict to configurations satisfying
$\tilde\m_{\msbm{C}}=0$, which can easily be 
implemented supersymmetrically, we have
\eqn\matrit{\eqalign{
\int_{\overline \CV_N^{1,1}}&
d^{N^2}\!t_0
\left(\prod_{i=1}^{4}
d^{N^2}\!t_i\;d^{N^2}\!t_i^*\;
d^{N^2}\!\p_i\;d^{N^2}\!\p_i^*\right)
\cr
&\times e^{
-\Fr{1}{2}
\tr\left( t_0\tilde\m_{\msbm{R}} 
+m(t_3t_3^*-t_4t_4^*)
+\vr^\pr(t_1t_1^*+t_3t_3^*)
+\vr^\ppr(t_2t_2^*+t_4t_4^*)
-i\sum_j\p_j\p_j^*\right)}
}
}
Integration over $t_0$ leads to an algebraic
equation of motion
$\m_{\msbm{R}}=0$. The integration
over $\p_i$ and $\p_i^*$ endows
$\overline\CV_N^{1,1}$ with the symplectic
measure. Then the remaining integration
is precisely the Duistermatt-Heckmann (DH) integration
formula for the $S^1_m\times S^1_{\vr^\pr}\times
S^1_{\vr^\ppr}$ actions \DH.
So the integration is localized to the fixed points
and can be evaluated exactly.
Note as one of fixed point locus of $S^1_m$
action we get the completed moduli space of
certain framed instantons.
 
For the particular integral \matrit\ we have
a natural prescription for resolving
the singularities.
Note that $\bs t_0 =\bbs t_0 =0$. We can
form an obvious observable, for example,
$-\Fr{1}{4}\tr t_0^2$.
Now we define a new matrix integral
\eqn\matrit{\eqalign{
\int_{\overline \CV_N^{1,1}}&
d^{N^2}\!t_0
\left(\prod_{i=1}^{4}
d^{N^2}\!t_i\;d^{N^2}\!t_i^*\;
d^{N^2}\!\p_i\;d^{N^2}\!\p_i^*\right)
\cr
&\times e^{
-\Fr{1}{2}
\tr\left(t_0\tilde\m_{\msbm{R}} 
+m(t_3t_3^*-t_4t_4^*)
+\vr^\pr(t_1t_1^*+t_3t_3^*)
+\vr^\ppr(t_2t_2^*+t_4t_4^*)
-i\sum_j\p_j\p_j^*\right)
-\Fr{\e}{4}\tr t_0^2}.
}
}
The algebraic equations of
of above action functional are given by
\eqn\resolp{
\tilde\m_{\msbm{R}} =- \e t_0,
\qquad\bbs\p_j =\bs\p^*_j=0.
}
Physically  introducing an $\e$ dependent term regularize the 
path integral. The norm squared of momentum map
$-\tr(\tilde\m_{\msbm{R}}^2)$ has no higher critical points. 
So introducing
$\Fr{\e}{4}\tr\left(t_0^2\right)$ does not introduce
any new fixed points. Instead it resolve the
singularities of the fixed point locus of
$S_m$ action. Thus we have another resolution
of singularity by the perturbation;
\eqn\aresol{
\tilde\m_{\msbm{C}} =0,
\qquad
\tilde\m_{\msbm{R}} =- \e t_0.
}
Now Witten's fixed point theorem ensures that
the matrix integral \matrit\ is localized
on the solutions of \aresol\ together
with the fixed points of supersymmetry $\bs$ and
$\bbs$ in \uaab. Clearly, the fixed points
are identical to the fixed point of the
torus action \asone. If we consider 
a fixed point of the $S^1_m$ action only
as example (2), we have
\eqn\xxapp{\left\{\eqalign{
&[z_1, w_1] + \s\pi=0,\cr
&[z_1,z_1^*] +[w_1,w_1^*] + \s\s^* 
-\pi^*\pi= -\e\a_{\msbm{R}}I_k,
}\right.\!\!\!\!\!\!\quad
\left\{\eqalign{
&[z_2, w_2] -\pi\s=0,\cr
&[z_2,z_2^*] +[w_2,w_2^*] 
 +\pi\pi^*-\s^*\s=\e\Fr{k\a_{\msbm{R}}}{n} I_n,\cr
}\right.
}
with $(1+k/n)\a= m$.
This is the real perturbation related
with the complex perturbation \xxap. 
The solution space of the first set of equation in \xxapp\
is
\eqn\resolb{
\overline\CM^{\e_{\msbm{R}}\a}_{n,k}=
\m_{\msbm{R}}^{-1}(-\e_{\msbm{R}}\a I_k)\cap 
\m_{\msbm{C}}^{-1}(0)
/U(k).
}
In this case the resolution of the singularities
occurs in the framework of the
variation of the symplectic quotients \GS.

We note that the above integral is  of the
form of the $U(N)$ equivariant integration
of Witten \WittenT.
Jefferey and Kirwan proved a version of
Witten's non-Abelian localization such
that the integral is localized to the
fixed points of the action of the maximal torus
of $U(N)$ \JK. Together with the DH type
localization, the integral \matrit\ is localized
on the fixed points of $T^N\times S^1_m\times 
S^1_{\vr^\pr}\times S^1_{\vr^\ppr}$ actions. 
In principle, the integral \matrit\ should be
exactly computable.

One can also do a Morse theory to evaluate
the Poincar\'e polynomial of $\overline\CN_N^{t_0}$.
Physically, this amounts to study the balanced
topological theory \DM. The contribution
of a particular fixed point locus of the $S^{1}_m$
action is identical to the partition function
of ${\cal N}=4$ super-Yang-Mills theory in four-dimensions
with some gauge group and instanton
number.  Surprisingly, such a theory is closely
related to  the dimensional reduction of ${\cal N}=4$ 
super-Yang-Mills
theory all the way down to zero-dimension.
The large $N$ limit of such balanced theory
should be also compared with the matrix model
approaches \MAT\ for $M$ theory \MT. 

The details will appear in a forthcoming paper \Park.

\lin{Some speculative remarks}

It will be very interesting to examine the $N\rightarrow
\infty$ limit of \matrit. Then, the fixed point locus
of $S^1_m$ action contains an arbitrary
produc  of resolved and
completed moduli spaces of $SU(n)$
instantons with all ranks $n$ and all instanton
numbers. This should be compared with the field theoretical
approach where one fixes a gauge group
and sums over the contributions of all instanton
numbers. In the matrix integral \matrit, contribution
from a  particular fixed can be identified
with the Donaldson invariants or topological
correlation function of ${\cal N}=2$ super-Yang-Mills
theory in four-dimensions \Donald\Witten. 
Our matrix integral puts the rank of gauge group 
and the instanton number on equal footing and
considers all possible moduli spaces simultaneously.
Due to the original suggestion of t' Hooft, we expect
the physics becomes simplified or has new noble properties
by taking $n\rightarrow \infty$ limit \tHooft. 
However, the reciprocity,
which interchange $n$ with $k$ for the instanton
contributions, seems to suggest us to reexamine the
physics of $n\rightarrow \infty$ limit. The viewpoint
we developed here seems to suggest that we better
consider all values of rank $n$ and the instanton
number $k$ simultaneously. 
Presumably, the genuine stringy property of
QCD becomes more manifest in the $N\rightarrow \infty$
limit in the new setting 
than in the usual large $n$ limit.

Consider Yang-Mills instantons on an ALE space 
(an ALE gravitational instanton).
Let $\eufm{N}({\bf u},{\bf w})$ denote  
a completed moduli space of framed instantons,
classified by the Mukai vectors ${\bf u}$ and ${\bf v}$,
after resolving the singularities.
Nakajima constructed irreducible representations
of affine Lie algebra using certain homology
class of $\eufm{N}$ \Nakajima\NakajimaC. 
An important
thing in his construction is 
a Lagrangian subvariety
$\eufm{N}({\bf u},{\bf w})$ in the product
of two-moduli spaces,  
$\eufm{N}({\bf u+Ce}^i,{\bf w})\times
\eufm{N}({\bf u},{\bf w})$. 
This Lagrangian subvariety is closely
related with another Lagrangian subvarieties
in the moduli spaces $\eufm{N}({\bf u+Ce}^i,{\bf w})$
and $\eufm{N}({\bf u}^i,{\bf w})$.
The later subvarieties are defined by the
compact sets of gradient flows between
the critical points of Hamiltonian analogous
to \hamill. Note that we  get,
as a fixed point of $S^1_m$ action, various
product spaces of  completed moduli space of framed 
instantons on $\msbm{R}^4$ with resolved singularities.
So we automatically get the natural setting
for Nakajima's story. One may start from
examining the gradient flow between
the critical points of \hamill. 

We also like to point out that
the matrix integral \matrit\
with the invariant polynomial
$V(t_0) = \sum_{j=2}^N \Fr{\e_j}{2j!}\tr (t_0^j)$
has a great similarity with the random matrix
model (See, for example, a collection
of the original papers \BW).
It would be certainly possible
that the matrix integral enjoys a recursion
relation for different values of $N$.

\subsubsec{The original ADHM case} 

Before leaving this section, we will consider
the case of the original ADHM construction. 
This is relevant since such a case appears
as a particular fixed point of our model.

The  $S^1_{\vr^\pr}\times S^1_{\vr^\ppr}$ 
action \asone\ descends to the
fixed point locus of the $S^1_{m}$ action
as considered in Sect.~$3.2$. 
On the moduli space
$\overline\CM^{\a_\msbm{R}}_{n,k}$ described
$(\t_1,\t_1,a,b)$, for example in \xxap,
the action becomes
\eqn\descended{
(\t_1,\t_2,a,b) \rightarrow 
(e^{i\vr^\pr}\t_1,e^{i\vr^\ppr}\t_2,
e^{i\vr^\pr}a,e^{i\vr^\ppr}b).
}
This action preserve the induced K\"{a}her
potential $\tilde \k$ on $\overline\CM^{\a_\msbm{R}}_{n,k}$;
\eqn\ikah{
\tilde\k =-\Fr{1}{2}\tr(\t_1\t_1^* 
+ \t_2\t_2^* + a^*a +bb^*).
}
On the fixed point locus
$\overline\CM^{\a_\msbm{R}}_{n,k}$,
\uaab\ reduces to
\eqn\uaabc{
\eqalign{
&\bs \t_j = \c_j,\cr
&\bbs \t_j^*=\c^*_j,\cr
}\qquad
\eqalign{
&\bbs\c_j=-i[\t_0,\t_j] 
+ i\vr^\pr \t_j \d_{1j}
+ i\vr^\ppr \t_j \d_{2j}
,\cr
&\bs\c_j^*=-i[\t_0,\t_j^*] 
- i\vr^\pr \t_j^*\d_{1j}
- i\vr^\ppr \t_j^*\d_{2j},\cr
}
}
where $j=1,2$ and
\eqn\uaabd{
\eqalign{
&\bs a = \eufm{a},\cr
&\bbs a^*=\eufm{a}^*,\cr
&\bs b = \eufm{b},\cr
&\bbs b^*=\eufm{b}^*,\cr
}\qquad
\eqalign{
&\bbs\eufm{a}=+i a\t_0 
+ i\vr^\pr a 
,\cr
&\bs\eufm{a}^*=-i\t_0a^* 
- i\vr^\pr a^*,\cr
&\bbs\eufm{b}=-i\t_0 b
+ i\vr^\ppr b 
,\cr
&\bs\eufm{b}^*=+i b^*\t_0 
- i\vr^\ppr b^*,\cr
}
}
where $\t_0$ denotes the generator of 
$U(k)$ action
$(\t_1,\t_2,a,b) \rightarrow
(g \t_1 g^{-1},g \t_2 g^{-1},a g^{-1},g b)$
where $g \in U(k)$.
The equivariant K\"{a}hler form is given
by
\eqn\ewq{\eqalign{
-\Fr{i}{2}(\bs\bbs -\bbs\bs)\tilde \k
=
&-\Fr{1}{2}\tr\left(\t_0\m_{\msbm{R}}\right) +
-\Fr{1}{2}\tr\left(
\vr^\pr \t_1\t_1^* + \vr^\ppr \t_2\t_2^* +
\vr^\pr a^*a +\vr^\ppr b b^*
\right)
\cr
&
+i\tr\left(\sum_{j}\chi_j\chi_j^* 
+ \eufm{a}^*\eufm{a} 
+\eufm{b}\eufm{b}^*\right).
}
}  

Provided that the restriction to $\m_{\msbm{C}}=0$
is understood, we can use the equivariant
K\"{a}hler form \ewq\ as the action functional.
We define
\eqn\actionk{\eqalign{
S(\e,\vr^\pr,\vr^\ppr) = 
&\Fr{1}{2}\tr\left(\t_0\m_{\msbm{R}}\right) +
\Fr{1}{2}\tr\left(
\vr^\pr \t_1\t_1^* + \vr^\ppr \t_2\t_2^* +
\vr^\pr a^*a +\vr^\ppr b b^*
\right)
\cr
&
-i\tr\left(\sum_{j}\chi_j\chi_j^* 
+\eufm{a}^*\eufm{a} 
+\eufm{b}\eufm{b}^*\right)
-\Fr{\e}{4}\tr \t_0^2.
}
}
Here, $\t_0$ is the generator of $U(k)$ action
and the role of the $\e$ dependent term is the resolution
of the singularities in $\hat{\overline \CM}_{k,n}$;
the $\t_0$ equation of motion is
\eqn\jkjkjk{
\m_{\msbm{R}} = [\t_1,\t_1^*] +  [\t_2,\t_2^*] 
+  bb^* -a^*a =\e \t_0.
}
This gives us a constraint
\eqn\cofkd{
\tr(bb^* -a a^*) = \e tr(t_0).
}
The term  $i\tr\left(\sum_{j}\chi_j\chi_j^* 
- \eufm{a}^*\eufm{a} -\eufm{b}\eufm{b}^*\right)$ 
is the K\"{a}hler form of $\hat{\overline \CM}_{k,n}$.
The role of the $\e$ dependent term is the resolution
of the singularities in $\hat{\overline \CM}_{k,n}$.
The  momentum map of $S^1_{\vr^\pr}\times S^1_{\vr^\ppr}$
is given by
\eqn\demom{
-\Fr{1}{2}\tr(\vr^\pr \t_1\t_1^* + \vr^\ppr \t_2\t_2^* +
\vr^\pr a^*a +\vr^\ppr b b^*), 
}
which critical points are identical to
the fixed points of supersymmetry in \uaabc\uaabd.
The matrix integral with the action functional \actionk\
is localized on, after applying
the non-abelian localization, 
the fixed point of $T^k \times 
S^1_{\vr^\pr}\times S^1_{\vr^\ppr}$.

If we set $\vr^\pr =\vr^\ppr =0$, the above
action functional is closely related to the action
functional of holomorphic Yang-Mills theory \HYM;
\eqn\hyma{
I(\e) =-\Fr{i}{4\pi^2}\int \tr(i\w F_A)\wedge\o
-\Fr{1}{4\pi^2}\int \tr(\p\wedge\bar\p)\wedge\o
-\Fr{\e}{8\pi^2}\int \tr(\w^2),
}
where $\o$ denotes the K\"{a}hler form. 
Here $\phi$ is the generator of infinite
dimensional gauge symmetry $\CG$.
The second term 
$\Fr{1}{4\pi^2}\int \tr(\p\wedge\bar\p)\wedge\o$
is identical to the $L^2$ K\"{a}hler form of
$\CM_{n,k}$. Due to the isometry mentioned
earlier the two theories are equivalent.
In the field theoretical approach, however, we
do not know how to complete the metric in
the moduli space. The approach based on the
ADHM construction is more simpler 
and well-defined. Furthermore
we can also utilize the 
$S^1_{\vr^\pr}\times S^1_{\vr^\ppr}$
actions, which are invisible in the field
theory cousin.

We can also do Morse theory using the Hamiltonian
\demom. This amounts to considering the balanced
topological theory based on the ADHM data.
The resulting theory is equivalent to ${\cal N}=4$
super-Yang-Mills theory which suffer the similar
problems like holomorphic Yang-Mills theory.
The details will appear in \Park.

In the papers \NakajimaC, Nakajima 
studied the resolution of singularities
in the completed moduli space of framed instantons,
say $\overline\CM_{n,k}$, on $\msbm{C}^2$.
He rigorously showed that the perturbation  
leads to a resolution of the singularities of
\orign.
He also studied the $S^1_{\vr^\pr}\times 
S^1_{\vr^\ppr}$ actions \descended\
and determined the fixed points and the Morse
index. The
momentum map \demom\ turns a perfect Morse functional.
Moreover, the critical point set consists of finite
points \NakajimaP.
This amounts to providing an algorithm to compute the 
Poincar\'e polynomial of
$\overline\CM_{n,k}^{\a_{\msbm{R}}I_k}$. 
See \NakajimaB\ for a model
computation for the rank $n=1$ case (the rank $1$
torsion-free sheaves).
Here we recovered Nakajima's approach
in terms of natural field theoretical method.

\newsec{D-instantons, Reciprocity
and D-branes} 

We have developed enough material to come to
the main theme of this paper,  D-instantons.

To begin with, We  recall some 
properties of the D-instanton following Witten \WittenE.
The Type IIB string can have $(-1)$-branes or D-instantons.
The low energy effective theory of $N$ D-instantons
y are given by the dimensional reduction
of ${\cal N} =1$ $d=10$ super-Yang-Mills theory with
$U(N)$ gauge group all the way down to
zero-dimension. 
For the supersymmetric ground state
of the D-instantons are determined
by ten $U(N)$ matrices $X^i$ with
\eqn\caa{
[X_i,X_j]=0, \qquad i,j = 1,....,10.
} 
This is the absolute minimum of the bosonic 
part of the action
\eqn\cab{
I=\Fr{T^2}{2}\sum_{i < j}\tr[X^i,X^j]^2 + \ldots,
}  
Now for classical states of the unbroken
supersymmetry given by \caa, the matrices
can be  diagonalized simultaneously. The eigenvalues
of $X_i$ can be interpreted as the positions
of $N$ unordered (identical) points. Expanding
around the classical vacua, one interprets
the off-diagonal components as representing
strings joining D-instantons.
The classical moduli space of D-instantons
is $S^N(\msbm{R}^{10})$. The moduli space
is singular when more than two D-instantons
coincide. Such a singularity corresponds
to certain gauge symmetry enhancement.
As the $n$ of the D-instantons coincide
the gauge symmetry is enhanced to $U(n)$.

As pointed out by Witten \WittenE,
the above descriptions of D-instanton has
an intriguing similarity with the ADHM description of
the Yang-Mills instantons, although the governing
equations are different.
The situation is just like the ADHM 
description of the point-like instantons, which correspond
to completely degenerated instantons. 
In this section we will discuss relations between
D-instantons and Yang-Mills instantons using
the dimensional reduction of twisted ${\cal N}=4$ super-Yang-Mills
theory with gauge group $U(N)$ in $4$-dimension down
to $0$-dimension.

\subsec{D-instantons and Yang-Mills instantons}

We will consider supersymmetric ground states
of D-instantons described by the equations
\summ,\summb\ and \mmde.\foot{
We will use complex notation such that
$t_j = X_{2j-1} + iX_{2j}$ for $j=1,\ldots,5$.} 
We are interested in the case that half of
the supersymmetry breaks down after turning
on the bare mass $m$ in \summb.
We already studied the corresponding supersymmetric
ground states in the previous sections.
The classical vacua are classified by
the eigenvalues of the diagonalized matrix
$t_5$. Assume, for simplicity, the case
of a partition $N= k + n$ such that
$t_5 = diag(\a I_k, \b I_n)$, the example $(2)$ of Sect.~$3.3.1$.
Then
the diagonal elements of of $t_3$ and $t_4$ are
identically zero.
This corresponds
to the case that $k$ D-instantons are confined
in a  $2$-complex dimensional
hyperplane defined by 
$t_3=t_4=0, t_5 =\a$ and $n$ D-instantons
are confined in another hyperplane defined
by $t_3=t_4=0, t_5 =\b$.

If $|\a -\b|\neq m$\foot{
If $\a=\b$, we have 
$$
[t_1,t_2] =0,\qquad [t_1,t_1^*] +[t_2,t_2^*]=0,
$$
which is the ADHM description of $N$ point-like
instantons on $4$-dimensional hyperplane
defined by $t_3=t_4=0,t_5 =\a$.
}, the off-diagonal
elements of $t_3$ and $t_4$ also vanish identically.
The fixed point equation we get is
\eqn\jkl{
\eqalign{
&[z_1, w_1]=0,\cr
&[z_1,z_1^*] +[w_1,w_1^*] =0,
}\quad\eqalign{
&[z_2, w_2] =0,\cr
&[z_2,z_2^*] +[w_2,w_2^*] =0,\cr
}
}
where $z_1,w_1$ are $k\times k$ and 
$z_2,w_2$ are $n\times n$ matrices.
In terms of the ADHM description of
Yang-Mills instantons
we end up with $k$ and $n$ 
point-like instantons in the first and the second
hyperplanes, repectively. Their positions
in the two hyperplanes are given
by the eigenvalues of $(z_1,w_1)$
and $(z_2, w_2)$ respectively.

Now let $|\a-\b| = m$, then
the off-diagonal parts of $t_3$ and
$t_4$ appears as given by \waste. 
We have
\eqn\xxa{\left\{\eqalign{
&[z_1, w_1] + \s\pi=0,\cr
&[z_1,z_1^*] +[w_1,w_1^*] + \s\s^* 
-\pi^*\pi=0,
}\right.\quad
\left\{\eqalign{
&[z_2, w_2] -\pi\s=0,\cr
&[z_2,z_2^*] +[w_2,w_2^*] 
 +\pi\pi^*-\s^*\s=0,\cr
}\right.
}
where 
$\s$ is $k\times n$ and $\pi$ is $n\times k$ matrices.
We also have
\eqn\tryr{
\tr\s\pi=\tr\pi\s=0,\qquad
\tr\s\s^* =\tr\pi^*\pi,\qquad
\tr\s^*\s =\tr\pi\pi^*.
}
The first set of equations in \xxa\ corresponds
to the $k$ $SU(n)$ instantons in the first
hyperplane and the second one does 
the $n$ $SU(k)$ instantons
in the second hyperplane.
Note that the role of $\s$ and $\pi$ are exchanged
in the two sets of the equations in \xxa.

Now we consider the role of the off-diagonal
parts $\s$ and $\pi$ of $t_3$ and $t_4$,
respectively.
Here we should distinguish
the non-degenerate instantons from the
degenerate one.
First we consider the non-degenerate case so that
we do not have any point-like instantons.
Note the non-degeneracy condition means
that every matrix element of
the $n\times k$ matrices $\pi$ and $\s^*$
is non-zero. The matrices 
$\pi$ and $\s^*$
can naturally be  interpreted 
as representing oriented strings starting from the 
$k$ D-instantons in first 
hyperplane and ending on the $n$ D-instantons
in the second hyperplane.
The $k\times n$ matrices $\s$ and $\pi^*$  correspond
to strings starting from the D-instantons $n$
in the second hyperplane
and ending on the $k$ D-instantons in the 
first hyperplane.
Note also that the $\tr(\pi^*\pi)$ corresponds
to the size of $k$ $SU(n)$ YM instantons in
the first hyperplane. 
Now consider $k$ $SU(n)$ instantons in the first hyperplane. 
We regard 
the second diagonal elements of $(z_1,w_1)$
as the positions of $k$ D-instantons in
the first hyperplane
(the half of the trace is the center mass).
The off-diagonal elements of $(z_1,w_1)$  can be interpreted
as string joining the $k$ D-instantons
confined in the first hyperplane.\foot{ 
Of course, the above description is well defined
modulo $U(k)$ symmetry. Strictly speaking, the positions
of D-instantons (equivalently point-like instantons)
in the first hyperplane is not well-defined.
Only the center of mass given by
$(\Fr{1}{k}\tr z_1,  \Fr{1}{k}\tr w_1)$
is well-defined.}

Now we consider the degenerate instantons
in the first hyperplane. For example, 
we consider the case that $\ell=2$. 
Then the matrices $z_1$ and $w_1$ become
\eqn\raa{
\eqalign{
z_1=\left(\matrix{z_1^\pr & 0\cr 0& z_1^\ppr
}\right),\cr
w_1=\left(\matrix{w_1^\pr & 0\cr 0& w_1^\ppr
}\right),\cr
}\qquad
\eqalign{
z_1^\ppr=\left(\matrix{ a_1&0\cr
0&a_2}\right),\cr
w_1^\ppr=\left(\matrix{b_1&0
\cr0&b_2}\right),\cr
}
}
and, from \xxa, we have
\eqn\rab{
\s =\left(\matrix{\s^\pr \cr 0 \cr 0}\right),
\qquad
\pi = \left(\matrix{\pi^\pr & 0 & 0}\right),
}
where $z_1^\pr, w_1^\pr$ are $(k-2)\times (k-2)$
matrices, $\s^\pr$ and $\pi^\pr$ are $(k-2)\times n$
and $n\times (k-2)$ matrices, respectively.
{}From \xxa, we have 
\eqn\rac{
\eqalign{
&[z_1^\pr, w_1^\pr] + \s^\pr\pi^\pr=0,\cr
&[z_1^\pr,z_1^{\pr*}] +[w_1^\pr,w_1^{\pr*}] 
+ \s^\pr\s^{\pr*} 
-\pi^{\pr *}\pi^\pr=0.
}
}
Now we have two point-like instantons whose positions
are given by the eigenvalues $(a_1,b_1)$ and $(a_2,b_2)$
together with $k-2$ $SU(n)$ non-degenerated
instantons. The form of the matrices above
clearly show there are no-strings joining
the genuine point-like instantons with
the $SU(n)$ $k-2$ instantons. 
If we look at the second column of
the equation \xxa\ describing the instantons
on the second hyperplane, we find by
examining $\pi$
that they describe $SU(k-2)$, $n$ instantons.

A similar analysis can be repeated for more general
solutions. For a given partition $N=\sum k_{i}$
with $i= 1,\ldots,n_p$ we have $k_{i}$ D-instantons
confined in $i$-th $4$-dimensional hyperplane
defined by $t_3=t_4=0, t_5 = \a_i$.  

Up to now  we have considered the supersymmetric
ground states after breaking half of the
supersymmetry. Now we consider the perturbation
breaking one quater of the supersymmetry (Sect.~4.1.1).
This leads to resolution of 
singularities. The singularities occur
in the moduli space because we allow the instantons
degenerate. For example the system described
by \raa\ and \rab\ corresponds to the $2$nd
lower strata $\CM_{n,k-2}\times S^2(\msbm{C}^2)$
of the completed moduli space $\overline\CM_{n,k}$.
The $2$nd
lower strata $\CM_{n,k-2}\times S^2(\msbm{C}^2)$
has singularities when  positions of
two points coincides. 
One the hand, the perturbed system
does not allow solutions like \raa\ with $a_1=a_2$ and
$b_1=b_2$. Here it is more convenient
to consider the real perturbation \aresol. 
Then
we discover off-diagonal
components like 
\eqn\lkii{
z_1^\ppr=\left(\matrix{ a_1&\a\cr
0&a_1}\right),\qquad
w_1^\ppr=\left(\matrix{b_1&\b
\cr0&b_1}\right),
}
with the case $(\a,\b)=(0,0)$
excluded \LNN.
Mathematically such a perturbation resolves
the singularities in the completed
moduli space of instantons. 
Physically, 
we get a new massless degree
of freedom due to $(\a,\b)$.

So far we have demonstrated that our extended monads
interpolate between D-instantons and the Yang-Mills
instantons in very interesting way. 
We may  identify the
extended monads with  D-instantons
and  Yang-Mills instantons as its special
sub-sector.
Note that those complex $2$-dimensional hyperplanes,
where the Yang-Mills instantons are living, 
are derived notions.

\subsec{Monads, Space-time, Reciprocity,
D-branes and Universal Instantons; Some conjectural
discussions}

We saw that certain special configurations
of D-instantons can be identified with
Yang-Mills instantons. 
An important point is that the identification
has been made via the ADHM description,
which is the Fourier-Nahm transformation
of instantons on $\msbm{R}^4/H$ with $H=\{pt\}$. 
Some properties, the reciprocal
relation and the isometry,  of the ADHM description of
Yang-Mills instantons has an almost perfect analogy
with the Fourier transformation. 
Abusing terminology, we will refer
those properties collectively
as weak reciprocity. It is know
that weak reciprocity holds 
for Yang-Mills instantons with an arbitrary
$H$.
We may use the term strong reciprocity
if the dual connections defined by the ADHMN data
in the dual (momentum) space are also instantons.  
For general $H$ strong reciprocity
does not hold. A notable exception is the case of
a flat torus where the Nahm transformation leads
to instantons on the dual torus with the rank of gauge
group and the instanton number exchanged \Shenk\BB.

On the other hand,  strong reciprocity does
not hold for instantons on $\msbm{R}^4$. This
is essentially due to the non-compactness of $\msbm{R}^4$,
which requires special asymptotic condition (framing) at 
infinity for the existence of solutions.
If we remove such framing, strong reciprocity holds. 
Consider $k$ number of point-like instantons
(completely degenerated Yang-Mills instantons
with any gauge group) in $\msbm{C}^2$. 
The monad (ADHM)
description of such instantons is 
\eqn\waa{
\eqalign{
[\t_1,\t_2] =0,\cr
[\t_1,\t_1^*] +[\t_2,\t_2^*]=0,\cr
}}
where $\t_1$ and $\t_2$ are $U(k)$-valued matrix.
Let $g\in U(k)$, we have the adjoint $U(k)$ symmetry
$(\t_1,\t_2) \rightarrow (g\t_1 g^{-1},g\t_2 g^{-1})$.
One can form the dual connection $\tilde A$
with complex covariant derivatives $\tilde D_1$ and 
$\tilde D_2$ in the dual space $(\msbm{C}^2)^*$.
{}From \waa, we have
\eqn\wab{
\eqalign{
F^{0,2}_{\tilde A}&=[\tilde{D}_1,\tilde{D}_2] =0,\cr
\L F^{1,1}_{\tilde A}
&=[\tilde{D}_1,\tilde{D}_1^*] +[\tilde{D}_2,\tilde{D}_2^*]=0,
\cr
}}
which shows that the $U(k)$ connection
$\tilde A$ is self-dual in $(\msbm{C}^2)^*$.
Now we can view the ADHM description of the
$k$ point-like instantons as the dimensional reduction
of the above equation for $U(k)$ instantons
on the {\it  momentum space} down to zero-dimension.

We recall that the ADHM matrices $T_i$
are related with the coordinates $x_i$ of $\msbm{R}^4$
by \Nahm\CGO
\eqn\ntf{
T_i = -\int x_i\p^\dagger \p d^4 x,
}
where $\p$ denotes normalizable cokernel of Dirac
operator in the instanton background. 
For example \DK, to any solution of the ADHM equations for
$k$ instantons we can
associate  centre of mass in $\msbm{R}^4$
with coordinates $\{x_i = \Fr{1}{k}\tr T_i\}$.
When the matrices $T_i$ commute (for point-like instantons),
one can interpret them as space-time coordinates (positions
of point-like instantons).
It is clear that the ADHM description of 
points (point-like instantons) is highly redundant.
The off-diagonal elements of $T_i$ start to play
an important role when we want to resolve the
singularities in $S^{k}(\msbm{R}^4)$.
Mathematically, one interprets the moduli space
of rank $1$ torsion-free sheaves\foot{
The original moduli space of $k$ $U(1)$
instanton moduli space is empty.
However, the completed moduli space after
resolution of singularities is non-empty
and identical to length $k$ Hilbert scheme of points
on $\msbm{R}^4$.}
on $\msbm{R}^4$ as natural resolution of $S^{k}(\msbm{R}^4)$
\LNN. In the standard physical interpretation, 
a singularity is just a sign of
new new massless degree of freedom which can be
resolved by an extension, for example, as described by \lkii.
The most natural explanation of such a degree of
freedom can be found by replacing the point-like instantons
with D-instantons. A D-instanton has $U(1)$
degree of freedom due to open string which
becomes massless when two D-instantons coincide.
Thus, the ADHM description is naturally directing
us to find a new geometry relevant to D-branes.

In the previous section, 
the appearance of the genuine ADHM description of 
Yang-Mills instantons as a special sector of D-instantons 
has motivated us to interpret the off-diagonal components
of $T_i$, in its generalized form to $10$-matrices, 
as representing open strings.
As a summary, {\it one may interpret
the Nahm transformation \ntf\ as a
map which relates a gauge field theoretic
description in the usual geometry to a stringy
one in a non-commutative geometry}. 
These considerations naturally lead us to
conjecture of the existence of certain universal
instantons, which may be responsible
for the D-brane physics.

As we mentioned in the introduction, the Fourier-Nahm
transformation can be generalized to Yang-Mills
instantons on $\msbm{R}^4/H$ with any subgroup of isometries on $\msbm{R}^4$ 
\Nahm.
Note that the effective theory of $N$ D-instantons
is ${\cal N} =1$ super-Yang-Mills theory in $d=10$ with
$U(N)$ gauge group translation
invariant in all directions. In other words,
it is the super-Yang-Mills theory on $\msbm{R}^{*1,9}$
with isometry group $\msbm{R}^{*1,9}$. Now we want
to identify the isometry group $\msbm{R}^{*1,9}$
with the dual $H^*$ of the isometry group $H=\{pt\}$
of the dual space $\msbm{R}^{1,9}$.
In other words, we identify the $U(N)$ connections of
$D=10$ SYM theory as the $10$ matrix functions
of $\msbm{R}^{*1,9}/H^*$ obtained by a Fourier-Nahm
transformation of $N$ number of certain instantons 
in $\msbm{R}^{1,9}$ with isometry group $H=\{pt\}$. 
Note that the SYM theory
is embedded in the space of ADHM  data.  
Note also that the number of D-instantons
is exchanged with the rank of the gauge group. 
This naturally suggests
that the usual description of D-instantons
via Witten's non-commutative geometry
can be viewed as a Fourier-Nahm transformation
of $N$ point-like instantons in $\msbm{R}^{1,9}$ with
isometry group $H=\{pt\}$. 
The effective theory of $N$  D $p$-branes, $p > -1$
is ${\cal N} =1$ SYM theory in $\msbm{R}^{*1,9}$ with
isometry group $H^* = \msbm{R}^{0,9-p}$.
One may interpret this as a Fourier-Nahm  
transformation of $N$ instantons in $\msbm{R}^{1,9}$
with isometry group $H = \msbm{R}^{p+1}$. 
That is, the $N$ $p$-branes are described
by $10$ number of $U(N)$-matrix functions of 
$\msbm{R}^{*1,9}/H^*$.

The above suggestion leads us to
some general questions;

(1) What is the real space-time? 
One may
regard the
usual ten-dimensional space-time $R^{1,9}$ as
an auxiliary space. According to this view-point,
the arena of the string theory is the space of
ADHMN-like data (monads) which encodes information of
various D-branes and their excitations (strings) as well
as the interactions.
The other possible viewpoint is to regard the usual
space-time more fundamental. On that spacetime,
there may exist universal instantons and  string
theory can be understood as a way of solving
problems by after Fourier-Nahm-like transformation.
We may regard the conjectural existence of the universal
instantons and reciprocity as a  first
principle. This seems to suggest that it may be fruitable to regard
the both spaces as one object. 

(2) Note that \waa\ is a special limit of \abg, where
the reciprocity becomes obscure due to the framing
(the source term) at the infinity \CGO\Nahm.
In the ADHM construction, the
source term encodes the data of the gauge group.
This suggests an interesting possibility. 
The hypothetical instanton reciprocal to the
D-instanton may be the completely degenerate
(or the zero-size) limit of certain universal instantons.
In that case, we can expect a source term (or framing
at the infinity in $10$-dimensions)
in the RHS of \caa.  
Our discussions in the
previous sections suggest that the most natural
origin of such a source term is the extra dimensions.
This is an exciting possibility related with 
the $F$-theory of Vafa \VafaF. 
If this conjecture turns out to be valid, we 
may expect to have a natural explantion of the additional
$2$-dimensions which is responsible to the $SL(2,\msbm{Z})$
duality of type IIB strings.

(3) The analogy with the Nahm 
transformation of  instantons on the flat $4$-torus 
suggests that reciprocity may be identified with  
$T$-duality. Related suggestions have already been made  
in refs.~\DOM\DED.   
In this paper we concentrated on D-branes in flat space-time.
Recently, Taylor studied D-branes on a torus \Taylor.
It will be interesting to generalize our case 
to curved space.

(4) Recently, it has been suggested \MAT\ 
that $M$-theory \MT\ may be realized
as the  $N\rightarrow \infty$
limit of $p=0$ D-branes \ZB, related tp
$11$-dimensional supermembrane \WHN.
The obvious counterpart for $F$-theory side 
is the $N\rightarrow \infty$ limit of D-instantons.
According to reciprocity, the matrix model M-theory
may be regarded as a special case with isometry
group $H=\msbm{R}$.

\lref\BSV{
M.~Bershadsky, V.~Sadov and C.~Vafa,
{\it D-branes and topological field
theories}, \np{B463}{1996}{420}.
}
\lref\Mirror{
A.~Strominger, S.-T.~Yau and E.~Zaslow,
{\it Mirror symmetry is $T$-duality},
hep-th/9606040\semi
D.~R.~ Morrison,
{\it The geometry underlying mirror symmetry},
alg-geom/9608006\semi
K.~Becker, M.~Becker, D.R.~Morrison,
H.~Ooguri, Y.~Oz and Z.~Yin,
{\it Supersymmetric cycles in exceptional
holonomy manifolds}, hep-th/9608116.
}

Finally, we would like to point out some issues
concering the model we considered in the paper.
We used the set of equations \summ\mmde, which
dimensional reduction of the Vafa-Witten
equations \daa, as a model for D-instantons. 
The same set of equations can
be obtained by reducing a self-duality equation
in Calabi-Yau $4$-folds (or $\msbm{R}^{8}$) introduced 
by Donaldson
and Thomas \DT:
\eqn\sddt{
F^{0,2}_+ = 0, \qquad F.\o =0,
}
where $\o$ is the K\"{a}hler form.
Here, 
$F^{0,2}_+$ denote self-dual part of
$(0,2)$-component of curvature with respect
to the norm 
\eqn\dtn{
(\a^{0,2},\a^{0,2}) = |\a^{0,2}|^2\wedge \overline{\o^{0,4}},
}
where $\o^{0,4}$ denotes the non-degenerated
holomorphic $4$-form. The natural origin
of the Vafa-Witten type equations  in
$8$-dimensional manifolds
has been pointed out in \BSV.
It will be very interesting to study Fourier-Nahm 
transformations of \sddt. 
This may be a good place
to test the conjecture in (2) above.
The set of equations \summ\mmde\
may be appear as a completely degenerate limit
of ADHM like construction in $\msbm{R}^{8}$. 
It seems to be reasonable to
expect that Fourier-Nahm transformation of
\sddt\ has reciprocity relating mirror Calabi-Yau
$4$-folds. This conjecture is motivated by 
the quantum mirror symmetry
as $T$-duality of $T^4$ fibration \Mirror.

Clearly, there remain
many questions and unsolved problems requiring further
study.

\ack{
I am very grateful to  R.~Dijkgraaf 
and B.~J.~Schroers for  collaboration 
on a related subject and for useful discussions, and 
to B.~J.~Schroers for introducing 
me the ADHMN construction 
and reading  this manuscript. 
I would like to thank 
F.~A.~Bais, F.~Hacquebord and Q-Han Park
for useful discussions.
I am grateful to H.~Nakajima for sending
me his papers and the preliminary version
of lecture notes, for useful communications
and pointing out an error.
Special thanks to H.~Verlinde for many
useful discussions, suggestions and encouragement
throughout this project. 
This work is supported by a pioneer fund of
NWO.
}

\listrefs
\bye